\documentclass[journal,10pt]{IEEEtran}
\usepackage{graphicx}
\usepackage{caption}
\usepackage{subcaption}
\usepackage{mathtools}
\usepackage{bm,amsmath, amssymb}
\usepackage{amsmath}
\usepackage{amsfonts}
\usepackage {amssymb,amsmath}
\usepackage{cite}
\usepackage{epstopdf}
\usepackage{color}
\usepackage{microtype}
\usepackage{algorithm}                  
\usepackage{algorithmic}                
\DeclareGraphicsExtensions{.pdf,.png,.jpg}

\newcommand {\Define} {\stackrel {\Delta} {=}  }
\newcommand{\mya}{\mathrel{\overset{\makebox[0pt]{{\tiny(a)}}}{=}}}
\newcommand{\myb}{\mathrel{\overset{\makebox[0pt]{{\tiny(b)}}}{=}}}
\newcommand{\myc}{\mathrel{\overset{\makebox[0pt]{{\tiny(c)}}}{=}}}

\usepackage{cite}

\begin{document}
	\title{\vspace{-7mm} Low Complexity Channel Estimation for OTFS Modulation with Fractional Delay and Doppler}
	\author{\IEEEauthorblockN{Imran Ali Khan and Saif Khan Mohammed}
		\IEEEauthorblockA{ \thanks{Imran Ali Khan and and Saif Khan Mohammed (Email: saifkmohammed@gmail.com) are with the Department of Electrical Engineering, Indian Institute of Technology Delhi (IITD), New Delhi, India. Saif Khan Mohammed is also associated with the Bharti School of Telecommunication Technology and Management (BSTTM), IIT Delhi. This work is supported by the Prof. Kishan Gupta and Pramila Gupta Chair at IIT Delhi.}}
	}
	\maketitle
	
	\vspace{-17mm}
	\begin{abstract}       
	We consider the problem of accurate channel estimation for OTFS based systems with few transmit/receive antennas, where additional sparsity due to large number of antennas is not a possibility. For such
	systems the sparsity of the effective delay-Doppler (DD) domain channel is adversely affected in the presence of channel path delay and Doppler shifts which are non-integer multiples of the delay and Doppler
	domain resolution. The sparsity is also adversely affected when practical transmit and receive pulses are used. In this paper we propose a Modified Maximum Likelihood Channel Estimation (M-MLE) method for OTFS based systems which exploits the fine
	delay and Doppler domain resolution of the OTFS modulated signal to decouple the joint estimation of the channel parameters (i.e., channel gain, delay and Doppler shift) of all channel paths into separate estimation of the channel parameters
	for each path. We further observe that with fine delay and Doppler domain resolution, the received DD domain signal along a particular channel path can be written as a product of a delay domain term and a Doppler domain term where the delay domain term is primarily dependent on the delay of this path and the Doppler domain term is primarily dependent on the Doppler shift of this path. This allows us to propose another method termed as the two-step method (TSE), where the joint two-dimensional estimation of the delay and Doppler shift of a particular path in the M-MLE method is further decoupled into two separate one-dimensional estimation for the delay and for the Doppler shift of that path. Simulations reveal that the proposed methods (M-MLE and TSE) achieve better channel estimation accuracy at lower complexity when compared to other known methods for accurate OTFS channel estimation.                      
	\end{abstract}
	
	\begin{IEEEkeywords}
		OTFS, Channel Estimation, Doppler Spread, High Mobility, Low Complexity.
	\end{IEEEkeywords}
	\normalsize
	\section{Introduction}
	Next generation wireless communication technologies are expected to support reliable and high throughput communication even for very high mobility scenarios (e.g., high speed train, aircraft-to-ground communication, Unmanned Aerial Vehicle (UAV) communication systems) \cite{IMT2020}. However, fifth generation (5G) communication technologies are based on Orthogonal Frequency Division Multiplexing (OFDM), which suffers from high inter carrier interference (ICI) (and therefore performance degradation) due to the Doppler spread in high mobility scenarios \cite{NR5G}. Recently Orthogonal Time Frequency Space (OTFS) modulation has been proposed, which has been shown to be robust to mobility induced Doppler spread \cite{HadaniOTFS1, HadaniOTFS2, HadaniOTFS3}. This robustness towards Doppler
	spread is due to the fact that in OTFS modulation the information symbols are embedded in the delay-Doppler (DD) domain where the effective channel matrix is sparse which allows for joint demodulation of all information symbols. A derivation of OTFS modulation
	from first principles has been recently presented in \cite{SKM4}. 
	
	Low complexity detection of OTFS signals has been considered in \cite{channel,Emanuele2,lmpa,OTFSMP,mimootfs,MMSEOTFS,OTFSBayes}.
	However, the performance of joint detection of all information symbols is dependent on the availability of accurate channel estimates of the effective DD domain channel.
         The robustness of OTFS to mobility induced Doppler spread is primarily due to the sparseness of the DD domain channel matrix. The effective DD domain channel matrix is sparse
         only when the delay and Doppler domain resolutions are sufficiently good, i.e., the number of sub-divisions ($M$) along the delay domain and the number of sub-divisions ($N$) along the Doppler domain are both large.
         With large $(M,N)$ and fractional delay/Doppler spread\footnote{\footnotesize{Fractional delay and Doppler refers to a general scenario where the channel path delays are not integer multiples of the delay domain resolution and the path Doppler shifts are not
         integer multiple of the Doppler domain resolution.}}, the effective DD domain channel is a large $MN \times MN$ sparse matrix, which still has several non-zero entries that need to be estimated.
         
         Several methods have been proposed for estimating the effective DD domain channel matrix. In \cite{HadaniOTFS2}, an estimation method has been proposed based on the transmission of impulse in the DD domain as a pilot signal. The received DD domain
         signal can then be used to estimate the effective DD domain channel. This impulse based method has been extended to OTFS MIMO systems in \cite{mimootfs}. Later, in \cite{EmChEst} a method was proposed where the pilot impulse was transmitted
         along with information symbols in the same OTFS frame, in order to reduce the channel estimation overhead. As the DD domain channel matrix is sparse, several methods have been proposed where the channel estimation problem is formulated as a
         sparse recovery problem, which is solved using Orthogonal Matching Pursuit (OMP) \cite{Tropp}.
         
         OMP based OTFS channel estimation has been considered in \cite{ChEstOTFS1, Dshi2021, OMPEsti}. In \cite{ChEstOTFS1}, the channel sparsity of a downlink OTFS massive MIMO system in the 3D delay-Doppler-angle domain is considered and a  3D-structured OMP (3D-SOMP) algorithm is proposed for estimation of the downlink OTFS massive MIMO channel. The work in \cite{ChEstOTFS1} considers random pilots. This work has been extended in \cite{Dshi2021} where deterministic pilots have been used instead of random pilots. In \cite{OMPEsti}, an OMP based estimation has been proposed for multiuser uplink communication where both the user terminals (UTs) and the base station (BS) have single-antenna each.
         
         The OMP based methods achieve good estimation accuracy only when the effective channel is sparse in some domain. The level of channel sparsity in the DD domain is reduced when we consider non-ideal transmit and receive pulses and when the channel path delay and Doppler shifts are non-integer multiples of the delay and Doppler domain resolution. Additional sparsity is achieved in OTFS based massive MIMO systems where the signal received from different paths can also be differentiated in the angle domain, as has been considered in \cite{ChEstOTFS1, Dshi2021}. However, for single-antenna systems, the angle domain sparsity cannot be exploited and achieving good channel estimation accuracy is a challenge. In this paper, we consider this problem of low-complexity channel estimation in OTFS based single-antenna systems where additional sparsity due to large number of antennas at the BS is not a possibility.
         
               For OTFS based single-antenna systems (where there is no additional sparsity due to multiple antennas), recently a parametric channel estimation approach has been considered in \cite{Zhao2020, Fliu2021, sparse1}, where instead of estimating the elements of the DD domain channel matrix, the channel path parameters (i.e., path gain, path delay and Doppler shifts) are estimated from the received pilot signals. The effective DD domain channel matrix can then be reconstructed from the estimated channel path parameters. As the number of channel path parameters is usually much smaller than the number of significant energy elements of the effective DD domain channel matrix, these parameters can be estimated effectively (without the need for additional sparsity) based on the sparse Bayesian learning (SBL) method \cite{Tipping, BCS2008}. SBL based OTFS channel estimation has been considered in  \cite{Zhao2020, Fliu2021, sparse1}.    
             
             Although \cite{Zhao2020, Fliu2021} propose an SBL based OTFS channel estimation method, they consider the channel path delays to be integer multiples of the delay domain resolution which is not a practical assumption
             for realistic scenarios. The SBL based OTFS channel estimation in \cite{sparse1} considers non-integer delay and Doppler shifts and is shown to achieve good channel estimation accuracy in single-antenna systems. However, SBL
             based OTFS channel estimation methods have high complexity due to the requirement of inversion of large sized matrices.    
                     
	    OTFS modulation based systems are expected to have fine delay and Doppler domain resolution. This is because, the robustness of OTFS modulation to channel induced delay and Doppler shifts is primarily due to the joint demodulation of all DD domain information symbols, which is practically feasible only when the effective DD domain channel is sparse, which in turn is possible when the delay and Doppler domain resolution is good.  
	    In this paper, we exploit the fine delay and Doppler domain resolution of OTFS based systems to propose two low-complexity channel estimation methods for single-antenna systems, which are shown to acquire accurate channel estimates
	    with practical rectangular transmit and receive pulses in channel scenarios where the delay and Doppler shifts are non-integer multiples of the delay and Doppler domain resolution respectively.
	    The novel contributions of this paper are as follows.
	     \begin{enumerate}
	     \item In this paper we consider the parametric estimation of the channel path gain, delay and Doppler shifts based on the received OTFS signals when an impulse like DD domain pilot is transmitted. We consider joint maximum likelihood (ML) estimation of these parameters. In Section \ref{secChannelest} we observe that with fine delay and Doppler domain resolution, the joint ML estimation of all paths decouples into separate estimation of the path gain, delay and Doppler shift of each path.
	     Based on this observation we propose a low-complexity Modified Maximum Likelihood Estimation (M-MLE) method in Section \ref{secChannelest}.   
	     \item The proposed M-MLE method is an iterative method which in a given iteration estimates the channel path gain, delay and Doppler shift of the strongest channel path from the residual received signal in that iteration and then subtracts a reconstructed estimate of the contribution of this channel path from the residual received signal, resulting in the received signal for the next iteration.
	     \item 
	     In Section \ref{subsec3B} we further observe that the contribution of a given channel path to the received DD domain pilot signal can be expressed as the product of a delay domain term and a Doppler domain term, where the delay domain term depends primarily on the delay of that channel path and the Doppler domain term depends primarily on the Doppler shift of that path. This allows us to further decouple the joint estimation of the delay and Doppler shift of the strongest channel path in each iteration of the M-MLE method into separate single-dimensional estimation of the path delay and the path Doppler shift. This method, termed as the two-step method (TSE) has even lower complexity than the M-MLE method.  
	     \item In Section \ref{simsec1} we compare the Normalized Mean Square Error (NMSE) performance of the proposed channel estimation methods (M-MLE and TSE) with that of the Impulse based channel estimation method in \cite{EmChEst}, the OMP method proposed in \cite{mimootfs} and the SBL method proposed in \cite{sparse1}, for the time-varying wireless channel between an arriving aircraft and the ground station. Through simulations we observe that with fine delay and Doppler domain resolution, the proposed methods achieve better NMSE performance when compared to the other considered methods (Impulse, OMP and SBL). We also compare the uncoded Symbol Error Rate (SER) performance achieved with the proposed channel estimates to that achieved with the other considered methods. It is observed that the SER performance achieved with the proposed methods is better than that achieved with the other considered methods and is close to the SER performance achieved with perfect channel state information (CSI). Further, the proposed estimation methods do not require matrix inversion and have significantly lower complexity when compared to the complexity of OMP and SBL based methods.  
	     \end{enumerate}
	 Notations: The following notations are used: $v$, ${\bf v}$ and ${\bf V}$ represent a scalar, vector and matrix respectively; ${ v}[n]$ and ${ V}[m,n]$ represent the $n$-th and $(m,n)$-th element of ${\bf v}$ and ${\bf V}$  respectively. The sign $\odot$ represents the Hadamard product (element wise multiplication). ${\bf V}^H$, ${\bf V}^T$ and ${\bf V}^*$ denote the Hermitian transpose, transpose, and complex conjugate of ${\bf V}$ respectively. $vec\left( {\bf V}\right)$ is column-wise vectorization of matrix ${\bf V}$ and $invec_{M,N}\left( {\bf v}\right)$ is  invectorization of vector ${\bf v}$ into a $M \times N$ matrix by filling the matrix column-wise. Also, ${\bf V}[:,k]$ denotes the $k$-th column of the matrix ${\bf V}$. For integers $q$ and $M$, $[ q]_M$ denotes the smallest non-negative integer which is congruent to $q$ modulo $M$. Also, for any real number $x$, $\lfloor x \rfloor$ denotes the greatest
	integer less than or equal to $x$. 
	\section{System Model}
	We consider a single-user OTFS modulation based system where the transmitter and receiver have a single antenna each.
	In OTFS systems, information symbols are embedded in the delay-Doppler (DD) domain. The DD domain is $T$ seconds wide along delay domain and $\Delta f = {1}/{T}$ Hz wide along Doppler domain. The delay domain is divided into $M$ equal parts (each ${T}/{M}$ seconds wide) and the Doppler domain is divided into $N$ equal parts (each ${\Delta f}/{N}$ Hz wide). The combination of a division along the delay domain and a division along the Doppler domain is called a Delay Doppler Resource Element (DDRE). There are therefore $MN$ DDREs. The $(l,k)$-th DDRE consists of the interval $[{(2l-1)T}/{2M}, {(2l+1)T}/{2M} )$ along the delay domain and the interval $[{(2k-1)\Delta f}/{2N}, {(2k+1)\Delta f }/{2N} )$ along the Doppler domain. To the $(l,k)$-th DDRE we assign a unique point $\left( {lT}/{M}, {k\Delta f}/{N}\right)$, which lies at the centre of the DDRE. The collection of $MN$ centre points corresponding to all the $MN$ DDREs is then referred to as the Delay Doppler Grid (DDG) which is given by the set of points $\Lambda$, i.e.
	
	{
		\small
		\begin{eqnarray}
				\label{DDG}
				\Lambda  \Define \hspace{-1mm} \left\{ \hspace{-0.5mm} \left( \frac{lT}{M} , \frac{k\Delta f}{N} \right)
				\Big| \, \, l=0, 1, \cdots M-1, k=0, 1, \cdots N-1 \hspace{-0.5mm} \right\}.
		\end{eqnarray}
		\normalsize}
	Each DDRE can carry one information symbol. Let $x[l,k]$ denote the information symbol transmitted on the $(l,k)$-th DDRE, i.e., there are totally $MN$ information symbols.
	In OTFS modulation, DD domain information symbols $x[l,k] \,,\, l=0,1,\cdots, M-1 \,,\, k=0,1,\cdots, N-1$ are firstly converted to Time-Frequency (TF) symbols $X[m,n]\,,\, m=0,1,\cdots, M-1\,,\, n=0,1,\cdots, N-1$
	through the inverse Symplectic Finite Fourier Transform (ISFFT) \cite{HadaniOTFS1}, i.e.
	
	{\vspace{-4mm}
		\small
		\begin{eqnarray}
			\label{otfs1}
			X[m,n] & \hspace{-3mm} = \hspace{-3mm} & \frac{1}{{MN}}  \sum\limits_{k=0}^{N-1}\sum\limits_{l=0}^{M-1}  x[l,k] \, e^{-j 2 \pi \left(\frac{m l}{M} - \frac{n k}{N}   \right)}.
		\end{eqnarray}  
		\normalsize}
	For a given $T > 0 $ and $\Delta f = 1/T$, these TF symbols are then used to generate the time-domain (TD) transmit signal which is given by the Heisenberg transform \cite{HadaniOTFS1}, i.e.
	
	{\vspace{-4mm}
		\small
		\begin{eqnarray}
			\label{otfs2}
			x(t) & = &  \sum\limits_{m=0}^{M-1}\sum\limits_{n=0}^{N-1} X[m,n] \, g(t - nT)  \, e^{j 2 \pi m \Delta f (t - nT)}
		\end{eqnarray}
		\normalsize}
	where $g(\cdot)$ is the transmit pulse. When $g(t)$ is approximately time-limited to $[0 \,,\, T]$, the Heisenberg transform in (\ref{otfs2}) is similar to OFDM
	where $X[m,n]$ is the symbol transmitted on the $m$-th sub-carrier ($m=0,1,\cdots, M-1$) of the $n$-th OFDM symbol ($n =0,1,\cdots, N-1$). Each OFDM symbol is of duration $T$ and the sub-carrier spacing is $\Delta f$, i.e., each OTFS frame has duration $NT$ and occupies bandwidth $M \Delta f$. 
	
	The multi-path wireless channel between the transmitter and the receiver consists of $L$ paths, where the complex channel gain, delay and Doppler shift of the $i$-th path ($i=1,2,\cdots,L$) are denoted by
	$h_i$, $\tau_i$ ($0 < \tau_i < T$) and $\nu_i$ respectively. In this paper, we denote the vector of channel path gains, channel path delays and Doppler shifts by ${\bf h} \Define [h_1, h_2, \cdots h_L]^T$, ${\boldsymbol \tau} \Define \left(\tau_1, \tau_2, \cdots, \tau_L \right)^T$ and ${\boldsymbol \nu} \Define \left(\nu_1, \nu_2, \cdots, \nu_L \right)^T$ respectively. 
	The delay-Doppler channel is given by \cite{Bello} 
	\begin{eqnarray}
		\label{channel}
		h(\tau, \nu) & =  & \sum_{i=1}^{L} h_i  \, \delta(\tau - \tau_i)  \, \delta(\nu - \nu_i).
	\end{eqnarray}
	The channel delay and Doppler shift taps of the $i$-th path are $\tau_{i} = \frac{l_{\tau_i} + \iota_{\tau_i}}{M \Delta f}$ and $\nu_{i} = \frac{k_{\nu_i} + \kappa_{\nu_i}}{NT}$ where $l_{\tau_i}$ and $k_{\nu_i}$ are the integer delay and Doppler indices of the $i$-th channel path and  $\iota_{\tau_i} \in [-0.5, 0.5]$ and $\kappa_{\nu_i} \in [-0.5, 0.5]$ are the fractional delays and Doppler shifts respectively. 
	In each OTFS frame a cyclic prefix (CP) of length $\tau_{max} \Define \max_i \tau_i$ is included, i.e.,
	for $-\tau_{max} \leq t \leq 0$, $x(t) = x(t + NT)$.
	Then, with $x(t)$ as the transmitted signal,
	the received signal is given by \cite{Bello}
	
	{\vspace{-4mm}
		\small
		\begin{eqnarray}
			\label{zse1}
			y(t) & =  & \sum\limits_{i=1}^L h_i  \, x(t - \tau_i) \, e^{j 2 \pi \nu_i (t - \tau_i)} \, + \, n(t)
		\end{eqnarray}
		\normalsize}where $n(t)$ is additive white Gaussian noise (AWGN) with power spectral density $N_0$.
	Most prior work
	consider a two-step OTFS receiver which is compatible with OFDM receivers and where the received TD signal $y(t)$ is first converted to a discrete TF signal
	through the Wigner transform \cite{HadaniOTFS1}, i.e., ${\Tilde Y}[m,n] \Define \int_{-\infty}^{\infty} g(t - nT) y(t) e^{-j 2 \pi m \Delta f t} dt \,,\, m=0,1,\cdots, M-1, n=0,1,\cdots, N-1$.
	This discrete TF signal is then converted to a DD domain signal through the Symplectic Finite Fourier Transform (SFFT) i.e.,
	${\widehat x}[l',k'] \Define \sum\limits_{m=0}^{M-1}\sum\limits_{n=0}^{N-1} {\Tilde Y}[m,n] e^{j 2 \pi \left( \frac{m l'}{M} - \frac{n k'}{N}\right)}  \,,\, k'=0,1,\cdots,N-1 , l'=0,1,\cdots,M-1$ \cite{HadaniOTFS1}.
	In this paper we consider the rectangular transmit and receive pulse, which is given by
	
	{\vspace{-4mm}
		\small
		\begin{eqnarray}
			\label{zprf7}
			g(t) & = 
			\begin{cases}
				\frac{1}{\sqrt{T}} &, 0 \leq t < T \\
				0 &, \mbox{\small{otherwise}} \\
			\end{cases}.
		\end{eqnarray} 
	}\normalsize
	From (\ref{otfs1}), (\ref{otfs2}), (\ref{zse1}), (\ref{zprf7}) and the two-step receiver operations (i.e., Wigner transform and SFFT), it follows that
	
	{\vspace{-4mm}
		\small
		\begin{eqnarray}
			\label{xhattwostep}
			{\widehat x}[l',k'] & = & \sum\limits_{k=0}^{N-1}\sum\limits_{l=0}^{M-1} x[l,k] \, {\widehat h}[l',k',l,k]  \,\, + \,\, {\widehat n}[l',k']
		\end{eqnarray}
		\normalsize}where ${\widehat n}[l',k']$ are the DD domain noise samples and the expression of ${\widehat h}[l',k',l,k]$ is given by (\ref{hklotfs}) (see top of next page).
	In (\ref{hklotfs}), for any real $x$, $\mbox{\small{sinc}}(x) \Define \sin(\pi x)/(\pi x)$.  
	\begin{figure*}
		{\vspace{-11mm}
			\small
			\begin{eqnarray}
				\label{hklotfs}
				\,\, \hspace{5mm} {\widehat h}[l',k',l,k] & \hspace{-3mm} =  &  \hspace{-3mm} \sum\limits_{i=1}^L h_i e^{-j 2 \pi \frac{\nu_i}{\Delta f} \frac{\tau_i}{T}}   \underbrace{\left[  \frac{1}{N} \sum\limits_{n=0}^{N-1} \hspace{-1mm} e^{-j 2 \pi n \left( \frac{k' - k}{N} - \frac{\nu_i}{\Delta f}  \right)} \right]}_{\mbox{\tiny{Doppler domain term}}}   \underbrace{\left[\frac{1}{M} \sum\limits_{m=0}^{M-1} e^{j 2 \pi \frac{m}{M} \left( l' -l -M \tau_i \Delta f\right)} f_{\tau_i,\nu_i,k,l'}(m) \right]}_{\mbox{\tiny{Delay domain term}}} \nonumber \\
				f_{\tau_i,\nu_i,k,l'}(m) & \hspace{-3mm} \Define  &  \hspace{-3mm} \sum\limits_{p =-m}^{M-1-m}  \hspace{-1mm} e^{j 2 \pi \frac{pl'}{M}}  {\Bigg [} \left( 1 - \frac{\tau_i}{T} \right) e^{j \pi \left(1 + \frac{\tau_i}{T}  \right) \left( \frac{\nu_i}{\Delta f} - p \right)}   \mbox{\small{sinc}}\left( \left(1 - \frac{\tau_i}{T} \right) \left(\frac{\nu_i}{\Delta f} - p \right) \right) \nonumber \\       
				& \hspace{-3mm}  & \, \, \, \, \, \, \,\, \, \, \, \, \, \, \, \,\, \, \, \, \, \, \, \, \,\, \, \, \, \, \, \, \, \,\, \, \, \, \, \, \, \, \,\, \, \, \, \, \, \, \, \,\, \, \hspace{-3mm} + e^{-j 2 \pi \frac{k}{N}} \left(\frac{\tau_i}{T} \right) e^{j \pi \left(\frac{\tau_i}{T}  \right) \left( \frac{\nu_i}{\Delta f} - p \right)}   \mbox{\small{sinc}}\left( \left(\frac{\tau_i}{T} \right) \left(\frac{\nu_i}{\Delta f} - p \right) \right) {\Bigg ]}.
			\end{eqnarray}
			\normalsize}
	\end{figure*}
	Further, ${\widehat n}[l',k']$ are i.i.d. ${\mathcal C}{\mathcal N}(0, M N N_0)$.
	The DD domain input-output relationship in (\ref{xhattwostep}) can be expressed in terms of the vector of transmitted and received DD domain symbols i.e., ${\bf x}$ and ${\widehat {\bf x}}$ respectively
	(see (\ref{eqn17}) on top of next page), i.e.
		\begin{eqnarray}
			\label{matrixEq1}
			{\widehat {\bf x}} & = &   \sum_{q=1}^{NM} \left({\bf B}_q\left( {\boldsymbol \tau}, {\boldsymbol \nu} \right) {\bf h} \right) \, {x}[q] \, + \,  {\widehat {\bf n}}  \, = \, {\bf G} {\bf x} +   {\widehat {\bf n}}	
		\end{eqnarray} 
	\begin{figure*}
	{\vspace{-6mm}
		\begin{eqnarray}
			\label{eqn17}
			{\widehat {\bf x}} & = &  \left( {\widehat x}[0,0] , \cdots,   {\widehat x}[M-1,0] ,   {\widehat x}[0,1] , \cdots,   {\widehat x}[M-1,1] ,  \cdots,  {\widehat x}[0, N-1] , \cdots,   {\widehat x}[M-1, N-1]  \right)^T \nonumber, \\
			{ {\bf x}} & = &  \left( { x}[0,0] , \cdots,   { x}[M-1,0] ,   { x}[0,1] , \cdots,   { x}[M-1,1] ,  \cdots,  { x}[0, N-1] , \cdots,   { x}[M-1,N-1]  \right)^T \nonumber, \\
			{\widehat {\bf n}} & = &  \left( {\widehat n}[0,0] , \cdots,   {\widehat n}[M-1, 0] ,   {\widehat n}[0, 1] , \cdots,   {\widehat n}[M-1, 1] ,  \cdots,  {\widehat n}[0, N-1] , \cdots,   {\widehat n}[M-1, N-1]  \right)^T.
		\end{eqnarray}}
		{\vspace{-3mm}
		\begin{eqnarray*}
		\hline
		\end{eqnarray*}}
	\end{figure*}where $x[q]$ is the $q$-th element of the DD domain information symbol vector ${\bf x}$ (see (\ref{eqn17}) on top of next page) and ${\bf B}_q\left( {\boldsymbol \tau}, {\boldsymbol \nu} \right) \in {\mathbb C}^{NM \times L}$. The element of ${\bf B}_q\left( {\boldsymbol \tau}, {\boldsymbol \nu} \right)$ in its $(k'M + l' + 1)$-th row and $i$-th column is denoted by $b_{q,i,l',k'}$ which is given by
		
	{\small
	\vspace{-4mm}
		\begin{eqnarray}
			\label{apikleqn}
			b_{q,i,l',k'} & \hspace{-3mm} = &  \hspace{-3mm} e^{-j 2 \pi \frac{\nu_i}{\Delta f} \frac{\tau_i}{T}} {\Bigg [} \frac{1}{N} \sum\limits_{n=0}^{N-1} \hspace{-1mm} e^{-j 2 \pi n {\big (} \frac{k' - \bigl \lfloor {\frac{ q-1}{M}} \bigr \rfloor }{N} - \frac{\nu_i}{\Delta f}  {\big )}} {\Bigg ]}  \nonumber \\
			& & \hspace{-20mm}  \left[\frac{1}{M} \sum\limits_{m=0}^{M-1} e^{j 2 \pi \frac{m}{M} \left( l' -[q-1]_M -M \tau_i \Delta f\right)} f_{\tau_i,\nu_i,\bigl \lfloor {\frac{ q-1}{M}} \bigr \rfloor,l'}(m) \right]
		\end{eqnarray}
		\normalsize}where $f_{\tau_i,\nu_i,k,l'}(\cdot)$ is given by (\ref{hklotfs}) (see top of this page). In (\ref{matrixEq1}), the effective DD domain channel matrix is denoted by ${\bf G}$ and its $q$-th column is ${\bf B}_q\left( {\boldsymbol \tau}, {\boldsymbol \nu} \right) {\bf h}$.
	
	\section{Channel Estimation}
	\label{secChannelest}
	In this section we propose two different low complexity methods to estimate the channel parameters $({\bf h}, {\boldsymbol \tau}, {\boldsymbol \nu})$. In the first method discussed in section III-A, we consider a modified Maximum Likelihood (ML) objective function which we optimize over a refined delay-Doppler grid in order to obtain accurate channel estimates at low complexity. In the second method discussed in section III-B, we propose a two-step estimation method, where in the first step we only estimate the channel paths delays, followed by the second step where we estimate the complex channel path gains and the Doppler shifts for each path detected in the first step.
	
	\subsection{Proposed Modified-ML Estimator (M-MLE)}
	\label{subsec3A}
	We consider a pilot only OTFS frame for the purpose of channel estimation. In this frame,
	a pilot is only transmitted on the $(l=l_p, k=k_p)$-th DDRE, i.e., the DD domain pilot symbols
	are given by
	
	{\vspace{-4mm}
		\small
		\begin{eqnarray}
			\label{Pilot1}
			x_p[l,k] & = 
			\begin{cases}
				\sqrt{M N E_p} &, l=l_p, k=k_p \\
				0 &, \mbox{\small{otherwise}} \\
			\end{cases}.
		\end{eqnarray} 
		\normalsize}where $E_p$ is the energy of the corresponding transmitted TD pilot signal.
	From (\ref{xhattwostep}) it follows that the received DD domain pilot signal is given by
	\begin{eqnarray}
		\label{rxpilot1}
		{\widehat x}_p[l',k'] & = &  \sqrt{MN E_p} \, {\widehat h}[l',k',l_p,k_p]  \,\, + \,\, {\widehat n}[l',k'].
	\end{eqnarray}

	We next organize ${\widehat x}_p[l',k'], k'=0,1,\cdots, N-1, l' = 0,1,\cdots, M-1$ into a vector ${\widehat {\bf x}_p}$ such that the $(k'M+l'+1)$-th element of this vector is ${\widehat { x}_p}[l',k']$.
	From (\ref{matrixEq1}) and (\ref{rxpilot1}) we get
	{		\begin{eqnarray}
			\label{rxpilotvec}
			{\widehat {\bf x}_p} & = &  {\bf A}( {\boldsymbol \tau}, {\boldsymbol \nu}) \, {\bf h} \, + \,  {\widehat {\bf n}} \nonumber \\
			& = & \sum_{i=1}^{L} h_i  \, {\bf a}(\tau_i, \nu_i)  \, +  \, {\widehat {\bf n}}
		\end{eqnarray}
	}where ${\bf A}( {\boldsymbol \tau}, {\boldsymbol \nu}) \Define \sqrt{MN E_p} \, {\bf B}_{k_p M + l_p +1}({\boldsymbol \tau}, {\boldsymbol \nu})$ (the element of ${\bf B}_{k_p M + l_p +1}({\boldsymbol \tau}, {\boldsymbol \nu})$ in its $k'M + l'+ 1$-th row and $i$-th column is given by $b_{q,i,l',k'}$ in (\ref{apikleqn}) with $q = (k_p M + l_p +1)$) and ${\bf a}({\tau_i}, {\nu_i}) \in {\mathbb C}^{MN \times 1}$ is the $i$-th column of ${\bf A}({\boldsymbol \tau}, {\boldsymbol \nu})$. Since the additive noise in (\ref{rxpilotvec}) is i.i.d. ${\mathcal C}{\mathcal N}(0, MNN_0)$,
	the ML estimate of the channel parameters $({\bf h},  {\boldsymbol \tau}, {\boldsymbol \nu})$ is
	given by
	\begin{eqnarray}
		\label{mlesteqn}
		\left( {\widehat {\bf h}},  {\widehat {\boldsymbol \tau}}, {\widehat {\boldsymbol \nu}} \right) & = & \arg \min_{({\bf h},  {\boldsymbol \tau}, {\boldsymbol \nu})} \,
		\left\Vert   {\widehat {\bf x}_p} \, - \,  {\bf A}( {\boldsymbol \tau}, {\boldsymbol \nu}) \, {\bf h}  \right\Vert^2.
	\end{eqnarray}
	From (\ref{mlesteqn}) it follows that, for a given $\left( {\boldsymbol \tau}, {\boldsymbol \nu} \right)$, the ML estimate of the vector of channel gains is given by
	\begin{eqnarray}
		\label{hesteqn1}
		{\widehat {\bf h}}\left( {\boldsymbol \tau}, {\boldsymbol \nu} \right) &  \hspace{-3mm}  \Define &    \hspace{-3mm}  \arg \min_{{\bf h}} \left\Vert   {\widehat {\bf x}_p} \, - \,  {\bf A}\left( {\boldsymbol \tau}, {\boldsymbol \nu} \right) \, {\bf h}  \right\Vert^2 \nonumber \\
		& \hspace{-3mm}  = &   \hspace{-3mm}  \left(  {\bf A}( {\boldsymbol \tau}, {\boldsymbol \nu})^H {\bf A}({\boldsymbol \tau}, {\boldsymbol \nu}) \right)^{-1}  \hspace{-2mm} {\bf A}( {\boldsymbol \tau}, {\boldsymbol \nu})^H \,  {\widehat {\bf x}_p}.
	\end{eqnarray}
	Using (\ref{hesteqn1}) in (\ref{mlesteqn}), the ML estimate of $\left( {\boldsymbol \tau}, {\boldsymbol \nu} \right)$
	is then given by
	
	{\small
	\vspace{-4mm}
		\begin{eqnarray}
			\label{mlesteqn3}
			({\widehat {\boldsymbol \tau}}, {\widehat {\boldsymbol \nu}}) & \hspace{-3mm}  = &   \hspace{-3mm}  \arg \max_{{\boldsymbol \tau}, {\boldsymbol \nu}}    \Big[  {\widehat {\bf x}_p}^H   {\bf A}( {\boldsymbol \tau}, {\boldsymbol \nu}) \left(  {\bf A}({\boldsymbol \tau}, {\boldsymbol \nu})^H {\bf A}({\boldsymbol \tau}, {\boldsymbol \nu}) \right)^{-1}   \nonumber \\
			&& \hspace{39mm} {\bf A}({\boldsymbol \tau}, {\boldsymbol \nu})^H \,  {\widehat {\bf x}_p}  \Big]. 
		\end{eqnarray} 
		\normalsize}Using (\ref{hesteqn1}) and (\ref{mlesteqn3}), the ML estimate of ${\bf h}$ is then given by
	\begin{eqnarray}
		\label{mlesteqn4}
		{\widehat {\bf h}} & = &  {\widehat {\bf h}}\left( {\widehat {\boldsymbol \tau}}, {\widehat {\boldsymbol \nu}} \right)  \nonumber \\
		& =  &  \hspace{-3mm}  \left(  {\bf A}\left( {\widehat {\boldsymbol \tau}}, {\widehat {\boldsymbol \nu}} \right)^H {\bf A}\left( {\widehat {\boldsymbol \tau}}, {\widehat {\boldsymbol \nu}} \right) \right)^{-1}  \hspace{-2mm} {\bf A}\left( {\widehat {\boldsymbol \tau}}, {\widehat {\boldsymbol \nu}} \right)^H \,  {\widehat {\bf x}_p}.
	\end{eqnarray}
	In practical systems based on OTFS modulation, $(M,N)$ are large due to which we have fine delay and Doppler domain resolution (i.e., small $1/(M \Delta f)$ and $1/(NT)$ respectively). With large $(M,N)$, the matrix ${\bf A}( {\boldsymbol \tau}, {\boldsymbol \nu})^H {\bf A}( {\boldsymbol \tau}, {\boldsymbol \nu})$ is diagonally dominant (i.e., ${\bf A}(  {\boldsymbol \tau}, {\boldsymbol \nu})^H {\bf A}( {\boldsymbol \tau}, {\boldsymbol \nu})$ is almost a scaled identity matrix). In Appendix \ref{appendixA} we have mathematically shown that as $N \rightarrow \infty$ (i.e. with Doppler domain resolution $(1/(NT)) \rightarrow 0$), any two columns of ${\bf A}( {\boldsymbol \tau}, {\boldsymbol \nu})$ (corresponding to two channel paths having different Doppler shifts) are asymptotically ($N \rightarrow \infty$) orthogonal to each other. From the analysis in Appendix \ref{appendixA}, it also follows that the two columns are almost orthogonal if $\frac{1}{NT} \ll \vert \nu_1 - \nu_2 \vert$ where $\nu_1$ and $\nu_2$ are the Doppler shifts of the two paths (i.e., the Doppler domain resolution is fine enough to resolve the two paths into different DDREs along the Doppler domain). Therefore, a good approximation to the exact ML estimator in (\ref{mlesteqn3}) is given by
	\begin{eqnarray}
		\label{MLapproximation}
		\left( {\widehat {\widehat {\boldsymbol \tau}}}, {\widehat {\widehat {\boldsymbol \nu}}} \right) & \hspace{-3mm}  = &   \hspace{-3mm}  \arg \max_{{\boldsymbol \tau}, {\boldsymbol \nu}}    \left[  {\widehat {\bf x}_p}^H   {\bf A}( {\boldsymbol \tau}, {\boldsymbol \nu})  {\bf A}( {\boldsymbol \tau}, {\boldsymbol \nu})^H \,  {\widehat {\bf x}_p}  \right]  \nonumber \\
		& \hspace{-3mm}  = &   \hspace{-3mm}  \arg \max_{{\boldsymbol \tau}, {\boldsymbol \nu}}     \left\Vert    {\bf A}( {\boldsymbol \tau}, {\boldsymbol \nu})^H \,  {\widehat {\bf x}_p}     \right\Vert^2  \nonumber \\
		& \hspace{-3mm}  = &   \hspace{-3mm}  \arg \max_{{\boldsymbol \tau}, {\boldsymbol \nu}}   \sum_{i=1}^L  \left\vert  {\bf a}({ \tau_i}, {\nu_i})^H  {\widehat {\bf x}_p}  \right\vert^2.
	\end{eqnarray}From (\ref{mlesteqn4}) and the diagonal dominance of ${\bf A}( {\widehat {\boldsymbol \tau}}, {\widehat {\boldsymbol \nu}})^H {\bf A}( {\widehat {\boldsymbol \tau}}, {\widehat {\boldsymbol \nu}})$, it also follows that a good approximation to the estimate of the vector of channel gains is given by
	\begin{eqnarray}
		\label{approxMLh}
		{\widehat {\widehat {\bf h}}} & \hspace{-3mm} = & \hspace{-3mm}  \frac{{\bf A}( {\widehat {\widehat {\boldsymbol \tau}}}, {\widehat {\widehat {\boldsymbol \nu}}})^H \,  {\widehat {\bf x}_p}}{MNE_p}.
	\end{eqnarray}
	Here we have also used the fact that for any $({\boldsymbol \tau}, {\boldsymbol \nu})$, the diagonal terms of ${\bf A}( { {\boldsymbol \tau}}, { {\boldsymbol \nu}})^H {\bf A}( {{\boldsymbol \tau}}, { {\boldsymbol \nu}})$ are equal to $MNE_p$, as explained in the following. From (\ref{rxpilotvec}) it follows that, the received DD domain signal vector along the $i$-th channel path is $h_i \, {\bf a}(\tau_i,\nu_i)$ and therefore the total received energy along this path is $\vert h_i \vert^2 \, {\bf a}(\tau_i,\nu_i)^H {\bf a}(\tau_i,\nu_i)$ where ${\bf a}(\tau_i,\nu_i)^H {\bf a}(\tau_i,\nu_i)$ is the $(i,i)$-th diagonal element of the matrix ${\bf A}( { {\boldsymbol \tau}}, { {\boldsymbol \nu}})^H {\bf A}( { {\boldsymbol \tau}}, { {\boldsymbol \nu}})$.
	Since the signal processing operations at the receiver (i.e., Wigner transform and SFFT) are inverse of the operations at the transmitter (i.e., ISFFT and Heisenberg transform), the total received energy along the $i$-th path
	is $\vert h_i \vert^2$ times the total transmitted pilot energy in the DD domain i.e., $MNE_p$ (see (\ref{Pilot1})). Hence, it follows that the diagonal entries of ${\bf A}( { {\boldsymbol \tau}}, { {\boldsymbol \nu}})^H {\bf A}( { {\boldsymbol \tau}}, { {\boldsymbol \nu}})$ are equal to $MNE_p$. 
	In (\ref{MLapproximation}) we observe that the objective function depends on the parameters $(\tau_i, \nu_i)$ of the $i$-th channel path, only through the $i$-th term $\left\vert  {\bf a}({ \tau_i}, {\nu_i})^H  {\widehat {\bf x}_p}  \right\vert^2$ in the summation in the R.H.S. of (\ref{MLapproximation}). Therefore, we propose to estimate the delay and Doppler shift of each path separately which significantly reduces the estimation complexity when compared to the joint estimation of the delay and Doppler shifts of all paths in (\ref{mlesteqn3}). 
	
\begin{figure}
	\centering
		\includegraphics[width=0.9\linewidth]{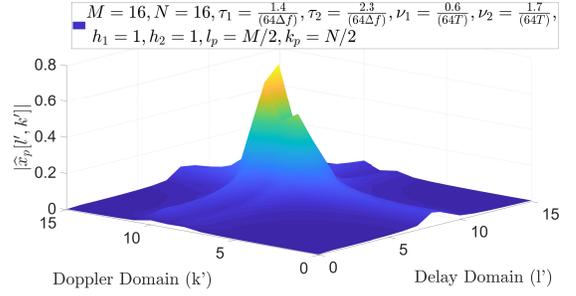}
		\caption{Received pilot signal in DD domain for two-path channel with coarse delay-Doppler resolution ($M = N = 16$).}
		\label{illusfig1}	
	\end{figure}

\begin{figure}
	\centering
		\includegraphics[width=0.9\linewidth]{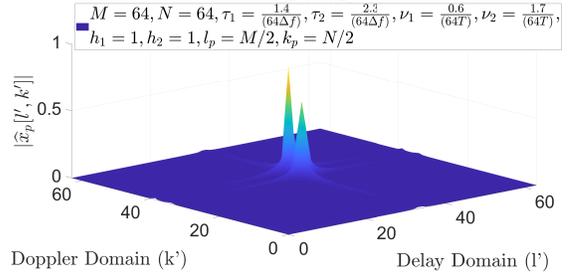}
		\caption{Received pilot signal in DD domain for two-path channel with fine delay-Doppler resolution ($M = N = 64$).}
		\label{illusfig2}	
	\end{figure}

	This separability of the joint estimation of multi-path delay and Doppler shifts is primarily due to the fine delay and Doppler domain resolution in OTFS based systems,
	due to which the pilot signal received along different channel paths are resolvable into different DDREs. We illustrate this through a simple example of a two-path channel. We consider two scenarios, i) coarse delay-Doppler resolution where due to small $M$ and $N$ the received pilot signal along the two paths are not resolvable into different DDREs (i.e., $|\tau_1 - \tau_2| < 1/(M \Delta f)$ and $| \nu_1 - \nu_2 |  < 1/(NT)$), and ii) fine delay-Doppler resolution where due to large $M$ or $N$ the received pilot signal along the two paths are resolvable into different DDREs. For this example, we plot the magnitude of the received DD domain pilot signal (i.e., $\vert \widehat{x}_p[l',k'] \vert$) for the coarse and fine delay-Doppler resolution scenarios in Fig.~\ref{illusfig1} and Fig.~\ref{illusfig2} respectively (the value of $\Delta f = 1/T$ and the values of the path gain, delay and Doppler shifts for both the paths, is the same for both these scenarios). From these figures it is observed that for the coarse delay-Doppler resolution scenario ($M = N = 16$) the received DD domain pilot signal along the two channel paths are not resolvable whereas with large $M=N=64$ (i.e., fine delay-Doppler resolution scenario) we observe two clearly separate and distinct peaks for the pilot signal received along the two channel paths.    

	 With the separability of the joint estimation of multi-path delay and Doppler shifts, the proposed estimate of the channel path gain, path delay and Doppler shift of the $i$-th path is given by
	\begin{eqnarray}
		\label{prpesteqn1}
		\left( {\widehat {\widehat { \tau_i}}}, {\widehat {\widehat { \nu_i}}} \right) & \hspace{-3mm}  = &   \hspace{-3mm}  \arg \max_{{ \tau_i}, { \nu_i}}  \left\vert  {\boldsymbol a}({ \tau_i}, { \nu_i})^H  {\widehat {\bf x}_p}  \right\vert^2,  \nonumber \\
		{\widehat {\widehat {h_i}}} & \hspace{-3mm} = & \hspace{-3mm}  \frac{{\boldsymbol a}(\widehat{ {\widehat \tau_i}}, \widehat{ {\widehat  \nu_i}})^H \,  {\widehat {\bf x}_p}}{M N E_p}.
	\end{eqnarray} 
	
	\normalsize
	Note that ${\boldsymbol a}({\tau_i}, {\nu_i}) \in {\mathbb C}^{MN \times 1}$ and its $(k'M + l' + 1)$-th element is equal to $b_{q,i,l',k'}$ in (\ref{apikleqn}) (with $q = k_pM+l_p+1$).
	We note that the proposed estimator in (\ref{prpesteqn1}) has much lower complexity than the ML estimator in (\ref{mlesteqn3}), since firstly the ML estimator in (\ref{mlesteqn3}) needs to perform the inverse of $\left(  {\bf A}( {\boldsymbol \tau}, {\boldsymbol \nu})^H {\bf A}( {\boldsymbol \tau}, {\boldsymbol \nu} \right))$
	and secondly that it is a $2L$-dimensional joint estimation of $2L$ parameters (i.e., $L$ path delays
	and $L$ Doppler shifts).
	
	Although the proposed estimator in (\ref{prpesteqn1}) has lower complexity than the ML estimator in (\ref{mlesteqn3}), it
	still needs to search over continuous valued path delay ($\tau$) and Doppler shift ($\nu$). We therefore propose to perform the maximization in (\ref{prpesteqn1}) over a refined version of the original DDG $\Lambda$ (defined in (\ref{DDG})). We refine the DD domain centre point corresponding to the $(l,k)$-th DDRE (i.e. $(\frac{lT}{M}, \frac{k\Delta f}{N})$) in the original DDG $\Lambda$ into $(2 \lfloor m_{\tau}/2 \rfloor +1)(2 \lfloor n_{\nu}/2 \rfloor +1)$ new DD domain points (refined points). The set of these new refined points in the $(l,k)$-th DDRE is given by (\ref{2Dgrid}) (see top of next page).
	\begin{figure*}
	\vspace{-7mm}
		\begin{eqnarray}
			{
				\label{2Dgrid}
				\Lambda^{(l, k)} \Define \left\{\left( \frac{lT}{M} + \frac{\gamma T}{m_{\tau} M}, \frac{k\Delta f}{N} + \frac{\chi \Delta f}{n_{\nu} N}\right)
				\Big| \, \, \gamma = -\bigl \lfloor \frac{m_{\tau}}{2} \bigr \rfloor \cdots, 0, \cdots \bigl \lfloor \frac{m_{\tau}}{2} \bigr \rfloor, \chi = -\bigl \lfloor \frac{n_{\nu}}{2} \bigr \rfloor \cdots, 0, \cdots \bigl \lfloor \frac{n_{\nu}}{2} \bigr \rfloor\right\}
			}
		\end{eqnarray}
				{\vspace{-3mm}
		\begin{eqnarray*}
		\hline
		\end{eqnarray*}}
	\end{figure*} For this refinement, we further sub-divide each DDRE into $(2 \lfloor m_{\tau}/2 \rfloor +1)$ equal sub-divisions along the delay domain and $(2 \lfloor n_{\nu}/2 \rfloor +1)$ equal sub-divisions along the Doppler domain.	
	
	In the proposed estimator, we estimate the delay, Doppler shift and complex channel gain of one path at a time, starting with the channel path along which highest energy is received in the DD domain. After estimating the path delay, Doppler shift and complex channel gain of this highest received energy path, we subtract its contribution from the received DD signal. We then repeat this process with the channel path along which the next highest energy is received. We therefore propose an iterative algorithm, where in each iteration, a path delay, Doppler shift, and channel gain is estimated and the contribution of the estimated path is cancelled from the received vector ${\widehat {\bf x}_p}$ to get the residual vector. 
	
	Let ${\widehat {\bf x}^{(t)}_p}$, denote the residual received DD domain vector at the start of the $t$-th iteration. Using (\ref{prpesteqn1}), the objective function for the $t$-th iteration is denoted by 
	\begin{eqnarray}
		\label{obj1}
		\Phi^{(t)}(\tau, \nu) & \Define &   \left\vert  {\boldsymbol a}({ \tau}, { \nu})^H  {\widehat {\bf x}^{(t)}_p}  \right\vert^2.
	\end{eqnarray}A listing of the proposed algorithm (M-MLE) is provided in Algorithm \ref{alg1list}. In the $t$-th iteration, we firstly compute the energy received in each DDRE. The received energy at the $(l,k)$-th DDRE is the squared absolute value of the DD domain symbol received in this DDRE i.e. $|{\widehat {\bf x}_p}^{(t)}[kM +l +1]|^2$ (see step $4$ in Algorithm \ref{alg1list}). Next, we find the DDRE having the highest received energy in the $t$-th iteration. Let the Doppler and delay domain index of this DDRE be denoted by $k^{(t)}$ and $l^{(t)}$ respectively (see steps $5$ to $7$ in Algorithm \ref{alg1list}). Next, we maximize the objective function $\Phi^{(t)}(\tau, \nu)$ (defined in (\ref{obj1})) for $(\tau, \nu)$ restricted to the refined DDG in the $(l^{(t)}-l_p, k^{(t)}-k_p)$-th  DDRE, i.e. $(\tau, \nu)$ in $\Lambda^{(l^{(t)}-l_p, k^{(t)}-k_p)}$ (see (\ref{2Dgrid})). This maximization is given by step 8 in Algorithm \ref{alg1list}. Let ${\widehat {\widehat { \tau}}}^{(t)}$ and ${\widehat {\widehat { \nu}}}^{(t)}$ denote the estimated path delay and Doppler shift of the path detected in the $t$-th iteration (see steps $8$, $9$ and $10$ in Algorithm \ref{alg1list}). Using (\ref{prpesteqn1}), the estimated complex channel gain for this detected path is then given by  ${\widehat {\widehat {h}}}^{(t)}  =  {\boldsymbol a}(\widehat{ {\widehat \tau}}^{(t)}, \widehat{ {\widehat  \nu}}^{(t)})^H \,  {\widehat {\bf x}_p}^{(t)}/ (MNE_p)$ (see step $11$ in Algorithm \ref{alg1list}). From (\ref{rxpilotvec}), we know that the contribution of the $i$-th channel path to the received DD domain vector ${\widehat {\bf x}}_p$ is $h_i  \, {\bf a}(\tau_i, \nu_i)$. Therefore in the proposed algorithm, before moving to the next iteration, we cancel the contribution of the path detected in this iteration, i.e., we subtract the contribution ${\widehat {\widehat { h}}}^{(t)} {\boldsymbol a}(\widehat{ {\widehat \tau}}^{(t)}, \widehat{ {\widehat  \nu}}^{(t)})$ of the estimated channel path from ${\widehat {\bf x}^{(t)}_p}$ resulting in the residual received DD domain vector ${\widehat {\bf x}_p}^{(t+1)}$ (see step $13$ in Algorithm \ref{alg1list}). Next we compute the total energy of this residual received signal and normalize it by the average received pilot signal power (see step $14$). The algorithm terminates if either, i) the maximum number of allowed iterations in Algorithm \ref{alg1list} (i.e. $T_{max}$) is reached or, ii) the difference between the normalized energy of the residual received vector in the current iteration and that of the previous iteration is less than a pre-determined threshold $\epsilon$ (subsequently referred to as the convergence tolerance parameter).
	\begin{algorithm}
		\caption{Proposed Modified ML Algorithm (M-MLE)}
		\label{alg1list}
		\begin{algorithmic}[1]
			\STATE \textbf{Input:} Received DD domain vector ${\widehat {\bf x}_p}$, Refinements $m_{\tau}$, $n_{\nu}$, Pilot location $(k_p, l_p)$
			
			\STATE \textbf{Initialization:} Convergence tolerance $\epsilon$, Counter $t=1$, Maximum iteration $T_{max}$, Residual received vector ${\widehat {\bf x}_p}^{(t)} = {{\widehat {\bf x}_p}}$, Normalized energy of residual vector $e^{(t)} = \frac{{\widehat {\bf x}_p}^{(t)H}{\widehat {\bf x}_p}^{(t)}}{\mbox{\tiny{Avg. Rx. Pilot Signal Power}}}$, Vector of estimated parameters  ${\widehat {\widehat {\bf h}}} = [ \, \, ]$, ${\widehat {\widehat {\boldsymbol\tau}}} = [ \, \, ]$, ${\widehat {\widehat {\boldsymbol \nu}}} = [ \, \, ]$.
			
			\REPEAT
			
			\STATE ${\boldsymbol{\mathcal{E}}}^{(t)} = {\widehat {\bf x}_p}^{(t)} \odot {\widehat {\bf x}_p}^{*(t)}$
			
			\STATE $q^{(t)} = \arg \max_{q} \mathcal{E}^{(t)}(q)$
			
			\STATE $k^{(t)} = \bigl \lfloor \frac{q^{(t)} -1}{M} \bigr \rfloor $ 
			
			\STATE $l^{(t)} = \left( (q^{(t)}-1) \hspace{-2mm} \mod M \right)$
			
			\STATE $\left( {\widehat {\widehat { \tau}}}^{(t)}, {\widehat {\widehat { \nu}}}^{(t)} \right)   =  \arg \max_{_{_{\hspace{-15mm} { (\tau, \nu)} \in  \Lambda^{( l^{(t)}-l_p, k^{(t)}-k_p)}}}} \hspace{-7mm} \Phi^{(t)}(\tau, \nu)  $ 
			
			\STATE ${\widehat {\widehat {\boldsymbol\tau}}} = [{\widehat {\widehat {\boldsymbol\tau}}},\, \, {\widehat {\widehat { \tau}}}^{(t)} ]$
			
			\STATE ${\widehat {\widehat {\boldsymbol\nu}}} = [{\widehat {\widehat {\boldsymbol\nu}}}, \, \, {\widehat {\widehat { \nu}}}^{(t)} ]$
			
			\STATE ${\widehat {\widehat {h}}}^{(t)}  =  {\boldsymbol a}(\widehat{ {\widehat \tau}}^{(t)}, \widehat{ {\widehat  \nu}}^{(t)})^H \,  {\widehat {\bf x}_p}^{(t)}/(M N E_p)$ 
			
			\STATE ${\widehat {\widehat {\boldsymbol h}}} = [{\widehat {\widehat {\boldsymbol h}}}, \, \, {\widehat {\widehat { h}}}^{(t)} ]$ 
			
			\STATE ${\widehat {\bf x}_p}^{(t+1)} = {\widehat {\bf x}_p}^{(t)} - {\widehat {\widehat { h}}}^{(t)} {\boldsymbol a}(\widehat{ {\widehat \tau}}^{(t)}, \widehat{ {\widehat  \nu}}^{(t)})$
			
			\STATE $e^{(t+1)} = \frac{ {\widehat {\bf x}_p}^{(t+1)H}{\widehat {\bf x}_p}^{(t+1)}}{\mbox{\tiny{Avg. Rx. Pilot Signal Power}}}$
			
			\STATE t = t+1
			
			\UNTIL{$t=T_{max}$ or $|e^{t}-e^{t-1}| \leq \epsilon$}
			
			\STATE \textbf{Output:} Estimated Parameters ${\widehat {\widehat {\bf h}}}$, ${\widehat {\widehat {\boldsymbol\tau}}}$, ${\widehat {\widehat {\boldsymbol \nu}}}$.
		\end{algorithmic}
	\end{algorithm}
	
	The maximum number of iterations $T_{max}$ is choosen to be larger than the maximum possible number of significant paths generally observed in the channel of interest. At the same time $T_{max}$ should not be too large due to complexity constraints. The quality of the proposed channel estimate depends on the convergence tolerance parameter $\epsilon$. If $\epsilon$ is large, then the proposed algorithm would run for fewer number of iterations due to which all paths may not be detected, resulting in inaccurate estimation of the effective DD domain channel matrix ${\bf G}$ and therefore a high NMSE (Normalized Mean Square Error) value. The NMSE is given by
	\begin{eqnarray}
	\label{nmseeqn}
	\mbox{\small{NMSE}} & \Define & \mathbb{E} \left[ \frac{\Vert {\bf G} - {\widehat {\widehat {\bf G}}} \Vert^2_F}{\Vert {\bf G} \Vert^2_F} \right]
	\end{eqnarray}where ${\widehat {\widehat {\bf G}}}$ is the reconstructed channel matrix whose $q$-th column is ${\bf B}_q\left( {\widehat {\widehat {\boldsymbol\tau}}}, {\widehat {\widehat {\boldsymbol \nu}}} \right) {\widehat {\widehat {\bf h}}}$ and $\Vert {\bf G} \Vert_F$ denotes the Frobenius norm of ${\bf G}$.
	
\begin{figure}
		\includegraphics[width=1.0\linewidth]{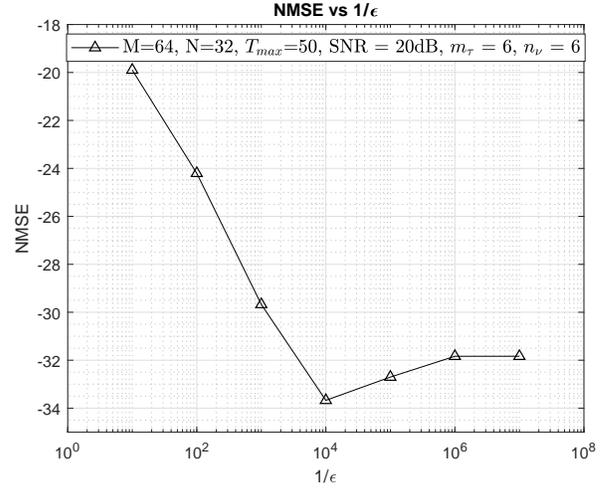}
		\caption{Impact of convergence tolerance parameter $\epsilon$ on the NMSE}
		\label{EpDecfig}	
	\end{figure}

	On the other hand a large value of $\epsilon$ may result in detection of false paths. Although at high SNR, these false paths are weaker than the detected true paths, they result in small degradation in the value of NMSE.
	In Fig.~\ref{EpDecfig} we have plotted NMSE vs $1/\epsilon$ for the proposed M-MLE algorithm. We have considered a channel model for the aircraft arrival scenario based on the model in \cite{HaarChannel}.
	The details of this channel model is described in the third paragraph of Section \ref{simsec1}.
	In Fig.~\ref{EpDecfig}, we consider five channel paths. The OTFS modulation parameters are $M=64$, $N=32$ and $T_{max} = 50$. In Fig. \ref{EpDecfig}, we observe that indeed as discussed above, the NMSE reduces when $\epsilon$ is reduced from $1$ to $10^{-4}$, after which further reduction in $\epsilon$ results in slight increase in the NMSE.  For a given pilot SNR (i.e., ratio of the received power of the time-domain pilot signal to the AWGN power at the receiver), we therefore choose $\epsilon$ in such a way that we achieve the minimum NMSE.
	
	From the listing of the proposed M-MLE algorithm, we observe that its total complexity is determined by step $8$ where we maximize the objective function $\Phi^{(t)}(\tau, \nu)$ on a refined grid.
	From (\ref{obj1}) we know that computing the objective function at a grid point $(\tau, \nu) \in \Lambda^{(l^{(t)} - l_p, k^{(t) } - k_p)}$ involves the computation of an inner product between the received DD domain vector ${\widehat {\bf x}_p}^{(t)}$ and ${\bf a}(\tau, \nu)$
	which has complexity $O(MN)$.    
	However, the delay and Doppler spread for each path is less than the maximum delay spread $\tau_{max}$ and Doppler spread $\nu_{max}$ respectively. Due to this, most of the received energy of the transmitted DD domain pilot symbol is localized in a small contiguous rectangular region of the DD domain, consisting of roughly $M_{\tau} = \lceil M \Delta f \tau_{max} \rceil +1$ DDREs along the delay domain and $N_{\nu} = 2 \lceil \nu_{max} NT \rceil +1$ DDREs along the Doppler domain. The order of complexity of step 8 and therefore that of each iteration of the proposed M-MLE algorithm is $O(m_{\tau} n_{\nu} M_{\tau} N_{\nu})$. The complexity of the other steps is also reduced due to the localization of the received energy of the DD domain pilot.
	Since there are roughly $O(P)$ number of iterations, the total complexity of the proposed M-MLE algorithm is $O(P m_{\tau} n_{\nu} M_{\tau} N_{\nu})$.
	Note that the proposed M-MLE algorithm does not require any matrix inversion and since its complexity scales only linearly with $ M_{\tau}$ and $N_{\nu}$,  this proposed method is applicable to channels where delay and Doppler spread is high.
	
	In the next section, we propose another estimation method (referred to as the Two-Step estimator (TSE))  which has lower complexity than the M-MLE method.
	In the proposed TSE method, for each channel path, we estimate its path delay in the first step, followed by the estimation of the Doppler shift and the channel path gain in the second step. This two step approach has lower complexity than the proposed M-MLE method as it does not involve the joint estimation of channel path delay and Doppler shift (see the two-dimensional optimization in step $8$ of the proposed M-MLE algorithm).
	
	\subsection{Proposed Two-Step Estimator (TSE)}	
	\label{subsec3B}
	In this proposed TSE method, the transmitted DD domain pilot signal is the same as that for the M-MLE method (see (\ref{Pilot1})) and therefore the received DD domain signal ${\widehat x}_p[\cdot, \cdot]$ is given by (\ref{rxpilot1}). We arrange ${\widehat x}_p[l', k']$, $l'=0, 1, \cdots M-1$, $k'=0, 1, \cdots N-1$ into a matrix $\widehat{\bf X}_p \in {\mathbb{C}}^{M \times N}$, such that the element of this matrix in its $(l'+1)$-th row and $(k'+1)$-th column is ${\widehat x}_p[l', k']$. In TSE, we firstly find the DDRE where the highest energy is received. This DDRE location is given by	
	\begin{eqnarray}
		\label{lklocation}
		(l'',k'') = \arg \max_{{(l,k)}} {\left \vert  {\widehat{ x}}_p[l,k]\right \vert}^{2}.
	\end{eqnarray}It is observed that the spreading of energy from the $(l'',k'')$-th DDRE to the other $(MN-1)$ DDREs is mostly localized to the DDREs $(l,k'')$ $(l=0,1, \cdots M-1)$ (i.e. entries in $(k''+1)$-th column of matrix ${\widehat{\bf X}}_p$) and the DDREs $(l'',k)$ $(k=0, 1, \cdots, N-1)$ (i.e. entries in the $(l''+1)$-th row of the matrix ${\widehat{\bf X}}_p$). This is evident from the expression of the received DD domain symbol in the $(l',k')$-th DDRE (i.e. ${\widehat{ x}}_p[l',k']$) in (\ref{rxpilot1}) where the noise-free term in the R.H.S. is the received pilot symbol multiplied with the effective channel gain ${\widehat h}[l',k',l_p,k_p]$. The energy of the received DD domain symbol in the $(l',k')$-th DDRE is therefore proportional to $|{\widehat h}[l',k',l_p,k_p]|^2$ whose expression from
	(\ref{hklotfs}) depends on the product of the delay domain term ${\left \vert \left[\frac{1}{M} \sum\limits_{m=0}^{M-1} e^{j 2 \pi \frac{m}{M} \left( l' -l_p -M \tau_i \Delta f\right)}  f_{\tau_i, \nu_i, k_p, l'}(m)\right] \right \vert}^2$ and the Doppler domain term  ${\left \vert \left[  \frac{1}{N} \sum\limits_{n=0}^{N-1} \hspace{-1mm} e^{-j 2 \pi n \left( \frac{k' - k_p}{N} - \frac{\nu_i}{\Delta f}  \right)} \right]    \right \vert}^2$ for the $i$-th channel path.
	Let $\tau'$ and $\nu'$ denote the delay and Doppler shift of the strongest channel path. Then, as the highest energy is received in the $(l'',k'')$-th DDRE, it is expected that with high probability $(l'' -l_p -M\tau'\Delta f) \in [-0.5,0.5]$ and $(k'' - k_p -\nu'NT) \in [-0.5,0.5]$. 
	
	For any $(l,k'')$-th DDRE $(l =0,1, \cdots, M-1)$, the energy of the DD domain signal received in this DDRE (i.e., $\vert {\widehat x}_p[l, k''] \vert^2$, see (\ref{rxpilot1})), depends on $\vert {\widehat h}[l,k'',l_p,k_p] \vert^2$ which in turn depends on the Doppler domain term $\left[  \frac{1}{N} \sum\limits_{n=0}^{N-1} \hspace{-1mm} e^{-j 2 \pi n \left( \frac{k'' - k_p}{N} - \frac{\nu'}{\Delta f}  \right)} \right]$ (as the expression of ${\widehat h}[l,k'',l_p,k_p]$ would contain contribution from the strongest channel path having delay and Doppler shift $\tau'$ and $\nu'$ respectively, see (\ref{hklotfs})).
	From the discussion in the previous paragraph, we know that with high probability $(k'' - k_p -\nu'NT) \in [-0.5,0.5]$ and therefore the Doppler domain term $\frac{1}{N} \sum\limits_{n=0}^{N-1} \hspace{-1mm} e^{-j 2 \pi n \left( \frac{k'' - k_p}{N} - \frac{\nu'}{\Delta f}  \right)} = \frac{1}{N} \sum\limits_{n=0}^{N-1} \hspace{-1mm} e^{-j 2 \pi n \left( \frac{k'' - k_p - \nu' N T} {N}  \right)} $ will have significant value due to which $\vert {\widehat x}_p[l, k''] \vert^2$ will have significant value for all $l=0,1,\cdots, M-1$. Similarly, for any $(k = 0, 1, \cdots, N-1)$, the energy received in the
	$(l'',k)$-th DDREs will also be significant.
		\begin{figure}
	\centering
		\includegraphics[width=0.9\linewidth, height=0.6\linewidth]{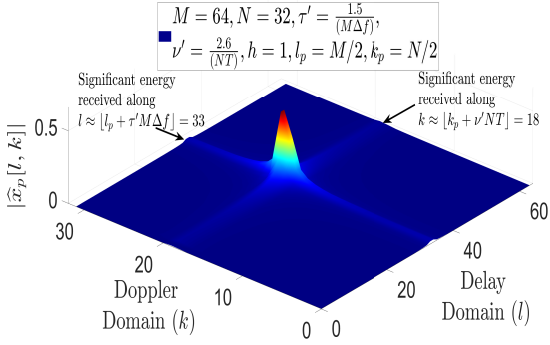}
		\caption{Received DD domain pilot signal for a single channel path having delay $\tau'$ and Doppler shift $\nu'$ ($M = 64, N = 32$).}
		\label{illusfig4}	
	\end{figure}
	This is illustrated in Fig.~\ref{illusfig4} for a single channel path (having delay $\tau'$ and Doppler shift $\nu'$), where significant pilot energy is received in the DDREs along the two light coloured lines and the maximum energy is received at the intersection of these two lines at the $(l'',k'')$-th DDRE ($l'' \approx \lfloor l_p + \tau' M \Delta f \rfloor  = 33$, $k'' \approx \lfloor k_p + \nu' N T \rfloor = 18$).
	    	
	In (\ref{hklotfs}) we also note that for each channel path, the path delay affects the effective DD domain channel gain primarily through the corresponding delay domain term and the path Doppler shift affects the effective DD domain channel gain primarily through the Doppler domain term.
	As discussed above, due to the strongest channel path with delay and Doppler shift $\tau'$ and $\nu'$ respectively, the Doppler domain term $\left[  \frac{1}{N} \sum\limits_{n=0}^{N-1} \hspace{-1mm} e^{-j 2 \pi n \left( \frac{k'' - k_p}{N} - \frac{\nu'}{\Delta f}  \right)} \right]$ appears in the expression of the channel gain of all the $(l, k'')$-th DDREs, $l=0,1,\cdots, M-1$. Similarly, the delay domain term $\left[\frac{1}{M} \sum\limits_{m=0}^{M-1} e^{j 2 \pi \frac{m}{M} \left( l'' -l_p -M \tau' \Delta f\right)}  f_{\tau', \nu', k_p, l''}(m) \right]$ appears in the expression of the channel gain of all the $(l'', k)$-th DDREs, $k=0,1,\cdots, N-1$.
	  
	As the delay domain term $\left[\frac{1}{M} \sum\limits_{m=0}^{M-1} e^{j 2 \pi \frac{m}{M} \left( l'' -l_p -M \tau' \Delta f\right)}  f_{\tau', \nu', k_p, l''}(m) \right]$ due to the strongest channel path is the same for the received DD domain signal in all the $(l'',k)$-th DDREs ($k=0,1,\cdots, k'', \cdots, N-1$) and the $(l'',k'')$-th DDRE (where the maximum energy has been received) is also in this list of DDREs, we can estimate the Doppler shift of the strongest channel path from the DD domain symbols received in all these $(l'',k)$-th DDREs $(k=0, 1, \cdots, N-1)$, i.e., from the entries in the $(l''+1)$-th row of the matrix ${\widehat{\bf X}}_p$.
Similarly, the delay of the strongest channel path can be estimated from the DD domain symbols received in the $(l,k'')$-th DDREs $(l=0, 1, \cdots, M-1)$, i.e., from the entries in the $(k''+1)$-th column of the matrix ${\widehat{\bf X}}_p$.   
		\vspace{2mm}
	\subsubsection{\underline{Delay Estimation}} 
	\label{subsubsec3B1}
	Let ${\widehat{\bf d}}_{k''}$ denote the $(k''+1)$-th column of ${\widehat {\bf X}}_p$ where $k''$ is given by (\ref{lklocation}) (see also paragraph above (\ref{lklocation})). Then from $(\ref{rxpilotvec})$ we have
			\begin{eqnarray}
				\label{rxpilotvecdelay}
				{\widehat{\bf d}}_{k''} & = & \sum_{i=1}^{L} h_i  {\bf a}_{k''}(\tau_i, \nu_i) + {\widehat {\bf n}_{k''}}
			\end{eqnarray}
		
	where ${\bf a}_{k''}(\tau_i, \nu_i) \in {\mathbb{C}}^{M \times 1}$ and its $(l+1)$-th element ($l=0, 1, \cdots, M-1$) is the $(k''M+l+1)$-th element of ${\bf a}(\tau_i, \nu_i)$.
	Similarly, ${\widehat {\bf n}}_{k''} \in {\mathbb{C}}^{M \times 1}$ and its $(l+1)$-th element ($l=0, 1, \cdots, M-1$) is the $(k''M+l+1)$-th element of ${\widehat {\bf n}}$.  
	Due to the fine delay-Doppler domain resolution of OTFS modulation, in the R.H.S. of $(\ref{rxpilotvecdelay})$ we separate the term corresponding to the strongest channel path, i.e.
		\begin{eqnarray}
		\label{rxpilotvecdelayappx1}
		{\widehat{\bf d}}_{k''} & = & h'  {\bf a}_{k''}(\tau', \nu') + {\Tilde {\bf n}_{k''}} \nonumber \\
		{\Tilde {\bf n}_{k''}} &  \hspace{-5mm} \Define &   \hspace{-6mm} \sum_{\substack{i=1  \\ i \, | \, (\tau_i, \nu_i) \ne (\tau', \nu')}}^{L} \hspace{-4mm} h_i  {\bf a}_{k''}(\tau_i, \nu_i) + {\widehat {\bf n}_{k''}}
	\end{eqnarray}where $\tau'$ and $\nu'$ are the delay and Doppler shift of the strongest channel path. As the Doppler domain term $\left[  \frac{1}{N} \sum\limits_{n=0}^{N-1} \hspace{-1mm} e^{-j 2 \pi n \left( \frac{k'' - k_p}{N} - \frac{\nu'}{\Delta f}  \right)} \right]$ corresponding to the strongest channel path is the same for all entries in the vector ${\bf a}_{k''}(\tau', \nu')$ and as observed earlier $(k'' -k_p -\nu'NT) \in [-0.5,0.5]$ with high probability, a good approximation to ${\bf a}_{k''}(\tau', \nu')$ in the R.H.S. of (\ref{rxpilotvecdelayappx1}) can be obtained by considering $(k'' -k_p -\nu'NT)/N = 0$ i.e., replacing $\nu'$ with $\frac{(k''-k_p)\Delta f}{N}$. Therefore
	\begin{eqnarray}
		\label{rxpilotvecdelayappx2}
		{\widehat{\bf d}}_{k''} & \approx & h'  {\bf a}_{k''}\left(\tau', \frac{(k''-k_p)\Delta f}{N} \right) + {\Tilde {\bf n}_{k''}}.
	\end{eqnarray}
	Note that ${\Tilde {\bf n}_{k''}}$ consists of the DD domain AWGN samples and the received DD domain symbols along the other weaker paths.
	Since $h'$ and ${\Tilde {\bf n}_{k''}}$ do not depend on the delay $\tau'$, we propose the following estimate of the delay of the strongest channel path, i.e.
	\begin{eqnarray}
		\label{MLapproximationdelay}
		{ {\widehat { \tau'}}} & \hspace{-3mm}  = &   \hspace{-3mm}  \arg \max_{{ \tau}}     \left\vert  {{\bf a}}_{k''}\left({ \tau, \frac{(k''-k_p)\Delta f}{N} }\right)^H  {\widehat {\bf d}_{k''}}  \right\vert^2
	\end{eqnarray}
	which is motivated by considering ${\Tilde {\bf n}_{k''}}$ to consists of contributions from several weak paths and can therefore be considered to have a statistical distribution like the zero mean complex Gaussian distribution. For practical implementation, we consider the maximization in (\ref{MLapproximationdelay}) over finite/discrete values of $\tau$ in the vicinity of the most likely delay of the strongest channel path i.e., $(l''-l_p)T/M$. That is, the estimate of the delay of the strongest channel path is then given by
		\begin{eqnarray}
		\label{MLapproximationdelay1}
		{\widehat {\widehat { \tau}'}} & \hspace{-3mm}  = &   \hspace{-3mm}  \arg \max_{{ \tau} \in \Lambda^{(l'' - l_p)}_{\tau}}     \left\vert  {{\bf a}}_{k''}\left({ \tau, \frac{(k''-k_p)\Delta f}{N} }\right)^H  {\widehat {\bf d}_{k''}}  \right\vert^2
	\end{eqnarray}
	where, for any $l \in (0,1,\cdots,  M-1)$, the set of discrete delay values $\Lambda^{(l)}_{\tau}$ is given by 

{\vspace{-4mm}	
	\small
	\begin{eqnarray} 
		{			
			\label{taugrid}
			\Lambda^{(l)}_{\tau} \Define \left\{\left( \frac{lT}{M} + \frac{\gamma T}{m_{\tau} M}\right)
			\Big| \, \, \gamma = -\bigl \lfloor \frac{m_{\tau}}{2} \bigr \rfloor \cdots, 0, \cdots \bigl \lfloor \frac{m_{\tau}}{2} \bigr \rfloor \right\}.
		}
	\end{eqnarray}}
\normalsize
	
	\subsubsection{\underline{Doppler shift and Channel gain Estimation}}
	\label{subsubsec3B2}
	After the estimation of the delay of the strongest channel path we estimate the Doppler shift of this strongest channel path from the $(l''+1)$-th row of the received $M \times N$ pilot matrix ${{\widehat {\bf X}}_p}$.  Let ${\widehat{\bf c}}_{l''} \in \mathbb{C}^{N \times 1}$ denote the $(l''+1)$-th column of ${{\widehat {\bf X}}_p}^T$ where $l''$ (given by (\ref{lklocation})) is the delay domain index of the DDRE where the highest energy is received. Then from $(\ref{rxpilotvec})$ we have
	\begin{eqnarray}
		\label{rxpilotvecDoppler}
		{\widehat{\bf c}}_{l''} & = & \sum_{i=1}^{L} h_i  {\bf b}_{l''}(\tau_i, \nu_i) + {\widehat {\bf n}_{l''}}
	\end{eqnarray}
	where ${\bf b}_{l''}(\tau_i, \nu_i) \in {\mathbb{C}}^{N \times 1}$ and its $(k+1)$-th element ($k=0, 1, \cdots, N-1$) is the $(kM+l''+1)$-th element of ${\bf a}(\tau_i, \nu_i)$.
	Similarly, ${\widehat {\bf n}}_{l''} \in {\mathbb{C}}^{N \times 1}$ and its $(k+1)$-th element ($k=0, 1, \cdots, N-1$) is the $(kM+l''+1)$-th element of ${\widehat {\bf n}}$.
	Just as for delay estimation in the previous sub-section, for Doppler estimation also, in (\ref{rxpilotvecDoppler}) we separate the term corresponding to the strongest channel path, i.e.
	\begin{eqnarray}
		\label{rxpilotvecDopplerappx1}
		{\widehat{\bf c}}_{l''} & = & h'  {\bf b}_{l''}(\tau', \nu') + {\Tilde {\bf n}_{l''}} \nonumber \\
		{\Tilde {\bf n}_{l''}} &  \hspace{-5mm} \Define &   \hspace{-6mm} \sum_{\substack{i=1  \\ i \, | \, (\tau_i, \nu_i) \ne (\tau', \nu')}}^{L} \hspace{-4mm} h_i  {\bf b}_{l''}(\tau_i, \nu_i) + {\widehat {\bf n}_{l''}}
	\end{eqnarray}where $\tau'$ and $\nu'$ are the delay and Doppler shift of the strongest channel path.
	With accurate estimation of the delay of the strongest channel path using (\ref{MLapproximationdelay1}), we can further approximate ${\widehat{\bf c}}_{l''}$ in (\ref{rxpilotvecDopplerappx1}) by replacing $\tau'$ in the R.H.S. of (\ref{rxpilotvecDopplerappx1}) by the proposed delay estimate ${\widehat {\widehat { \tau}'}}$, i.e.
	\begin{eqnarray}
		\label{rxpilotvecDopplerappx2}
		{\widehat{\bf c}}_{l''} & \approx & h'  {\bf b}_{l''}({\widehat {\widehat \tau'}}, \nu') + {\Tilde {\bf n}_{l''}}.
	\end{eqnarray}

	Since $h'$ and ${\Tilde {\bf n}_{l''}}$ do not depend on the Doppler shift $\nu'$, we propose the following estimate of the Doppler shift of the strongest channel path, i.e.
	\begin{eqnarray}
		\label{MLapproximationDoppler}
		{ {\widehat { \nu'}}} & \hspace{-3mm}  = &   \hspace{-3mm}  \arg \max_{{ \nu}}     \left\vert  { {\bf b}}_{l''}\left( {\widehat {\widehat \tau'}},  \nu \right)^H  {\widehat {\bf c}_{l''}}  \right\vert^2.
	\end{eqnarray}	
    For practical implementation of (\ref{MLapproximationDoppler}), we consider the optimization in (\ref{MLapproximationDoppler}) over discrete Doppler shift values in the vicinity of the most likely Doppler shift of the strongest channel path i.e. $(k''-k_p)\Delta f/N$. The proposed estimate of the Doppler shift of the strongest path is then given by
    \begin{eqnarray}
    	\label{MLapproximationDoppler1}
    	{\widehat {\widehat { \nu}'}} & \hspace{-3mm}  = &   \hspace{-3mm}  \arg \max_{{ \nu} \in \Lambda^{(k''  - k_p)}_{\nu}}     \left\vert  { {\bf b}}_{l''}({ {\widehat {\widehat {\tau}'}}, \nu})^H  {\widehat {\bf c}_{l''}}  \right\vert^2
    \end{eqnarray}
    where, for any $k \in (0,1,\cdots,  N-1)$, the set of discrete Doppler shift values $\Lambda^{(k)}_{\nu}$ is given by

		{\vspace{-4mm}
		\small
		\begin{eqnarray}
		{
			\label{Dopplergrid}
			\Lambda^{(k)}_{\nu} \Define \left\{\left(  \frac{k\Delta f}{N} + \frac{\chi \Delta f}{n_{\nu} N}\right)
			\Big| \, \, \chi = -\bigl \lfloor \frac{n_{\nu}}{2} \bigr \rfloor \cdots, 0, \cdots \bigl \lfloor \frac{n_{\nu}}{2} \bigr \rfloor\right\}.
		}
	\end{eqnarray}
	\normalsize}
	After estimating the delay and Doppler shift of the strongest channel path, its complex channel gain is estimated using (\ref{prpesteqn1}), i.e., the estimate of the complex channel gain of the strongest channel path is given by 
		\begin{eqnarray}
		\label{hestimate2}
		{\widehat {\widehat {h'}}} & \hspace{-3mm} = & \hspace{-3mm}  {\boldsymbol a}(\widehat{ {\widehat \tau'}}, \widehat{ {\widehat  \nu'}})^H \,  {\widehat {\bf x}_p} / (MN E_p)
	\end{eqnarray} 
	where the vector of received DD domain pilots (i.e. ${\widehat {\bf x}_p}$) is given by (\ref{rxpilotvec}).
	
	As we have seen above,  in the proposed TSE method, the estimation of the delay and Doppler shift of the strongest channel path is done seperately as two different single dimensional optimization (see (\ref{MLapproximationdelay1}) and (\ref{MLapproximationDoppler1})). Due to the single dimensional optimization, the overall complexity of TSE is lower than that of M-MLE where two dimensional optimization is performed to jointly estimate the delay and Doppler shift of the strongest channel path (see step $8$ in Algorithm \ref{alg1list}). We will discuss the overall complexity comparison between M-MLE and TSE later in this section.
	
	Just as in M-MLE, for TSE also we propose iterative estimation of the channel gain, delay and Doppler shifts of the channel paths. In each iteration, we estimate the delay, Doppler shifts and the complex channel gain of the strongest channel path and then using these estimates we subtract the reconstructed DD domain signal received along the strongest channel path from the received DD domain signal.
	\begin{algorithm}
		\caption{Proposed Two-Step Algorithm (TSE)}
		\label{alg2list}
		\begin{algorithmic}[1]
			\STATE \textbf{Input:} Matrix of received DD domain symbols ${\widehat{\bf X}}_p$, refinements $(m_{\tau}$, $n_{\nu})$ and pilot location $(k_p, l_p)$
			
			\STATE \textbf{Initialization:} Convergence tolerance $\epsilon$, Counter $t=1$, Maximum iteration $T_{max}$, $\, \, $  Residual received signal ${{\widehat{\bf X}}_p}^{(t)} = {\widehat{\bf X}}_p, \, \,$ Residual received vector ${\widehat {\bf x}_p}^{(t)} = vec \left( {\widehat {\bf X}_p}^{(t)}  \right)$, Normalized energy of residual received signal $e^{(t)} = \frac{{\widehat {\bf x}_{ p}}^{(t)H}{\widehat {\bf x}_{p}}^{(t)}}{\mbox{\tiny{Avg. Rx. Pilot Signal Power}}}$, Vector of estimated parameters  ${\widehat {\widehat {\bf h}}} = [ \, \, ]$, ${\widehat {\widehat {\boldsymbol\tau}}} = [ \, \, ]$, ${\widehat {\widehat {\boldsymbol \nu}}} = [ \, \, ]$.
			
			\REPEAT
			
			\STATE ${\boldsymbol{\mathcal{E}}}^{(t)} = {\widehat {\bf x}_p}^{(t)} \odot {\widehat {\bf x}_p}^{(t)*}$
			
			\STATE $q^{(t)} = \arg \max_{q} \mathcal{E}^{(t)}(q)$
			
			\STATE $k^{(t)} = \bigl \lfloor \frac{q^{(t)} -1}{M} \bigr \rfloor $ 
			
			\STATE $l^{(t)} = \left( (q^{(t)}-1) \hspace{-2mm} \mod M \right)$
			
			\STATE ${\widehat{\bf d}}_{_{{ {k}^{(t)} +1 } }} = {{\widehat{\bf X}}_p}^{(t)}(: ,{ { k } }^{(t)} +1) $
			
			\STATE ${\widehat {\widehat { \tau}}}^{(t)} =  \arg \hspace{-3mm} \max\limits_{{ \tau} \in \Lambda^{(l^{(t)}-l_p)}_{\tau}}     \left\vert  {\widehat {\bf a}}_{_{k^{(t)}+1}}({ \tau, \frac{( {k}^{(t)} - k_p)}{NT}})^H  {\widehat {\bf d}_{_{k^{(t)}+1}}}  \right\vert^2$ 
						
			\STATE ${\widehat {\widehat {\boldsymbol\tau}}} = [{\widehat {\widehat {\boldsymbol\tau}}},\, \, {\widehat {\widehat { \tau}}}^{(t)} ]$
			
			\STATE ${\widehat{\bf c}}_{{l^{(t)}+1}} = {{{\widehat{\bf X}}_p}^{(t)^T}}(:, l^{(t)} +1)$
			
			\STATE ${\widehat {\widehat { \nu}}}^{(t)}  =  \arg \hspace{-3mm} \max\limits_{{ \nu} \in \Lambda^{(k^{(t)}-k_p)}_{\nu}}     \left\vert  {\widehat {\bf b}}_{l^{(t)} +1}({ {\widehat {\widehat {\tau}}}^{(t)}, \nu})^H  {\widehat {\bf c}_{l^{(t)}+1}}  \right\vert^2$ 
			
			\STATE ${\widehat {\widehat {\boldsymbol\nu}}} = [{\widehat {\widehat {\boldsymbol\nu}}}, \, \, {\widehat {\widehat { \nu}}}^{(t)} ]$
			
			\STATE ${\widehat {\widehat {h}}}^{(t)} =   {\boldsymbol a}(\widehat{ {\widehat \tau}}^{(t)}, \widehat{ {\widehat  \nu}}^{(t)})^H \,  {\widehat {\bf x}^{(t)}_p}/ (M N E_p)$ 
						
			\STATE ${\widehat {\widehat {\boldsymbol h}}} = [{\widehat {\widehat {\boldsymbol h}}}, \, \, {\widehat {\widehat { h}}}^{(t)} ]$
			
			\STATE ${\widehat {\bf x}_p}^{(t+1)} = {\widehat {\bf x}_p}^{(t)} - {\widehat {\widehat { h}}}^{(t)} {\boldsymbol a}(\widehat{ {\widehat \tau}}^{(t)}, \widehat{ {\widehat  \nu}}^{(t)})$
			
			\STATE ${\widehat {\bf X}_p}^{(t+1)} = invec_{M,N} \left( {\widehat {\bf x}_p}^{(t+1)}  \right)$
			
			\STATE $e^{(t+1)} = \frac{ {\widehat {\bf x}_{ p}}^{(t+1)H}{\widehat {\bf x}_{ p}}^{(t+1)} } {\mbox{\tiny{Avg. Rx. Pilot Signal Power}}} $
			
			\STATE t = t+1
			
			\UNTIL{$t=T_{max}$ or $|e^{t}-e^{t-1}| \leq \epsilon$}
			
			\STATE \textbf{Output:} Estimated Parameters ${\widehat {\widehat {\bf h}}}$, ${\widehat {\widehat {\boldsymbol\tau}}}$, ${\widehat {\widehat {\boldsymbol \nu}}}$.
		\end{algorithmic}
	\end{algorithm}
		
	The listing of the proposed TSE method is provided in Algorithm \ref{alg2list}. In steps $4$ to $7$ we find the delay and Doppler index where the maximum energy is received in the DD domain. In the $t$-th iteration, these delay and Doppler domain indices are denoted by $l^{(t)}$  and $k^{(t)}$ respectively. In step $9$, using (\ref{MLapproximationdelay1}) we estimate the delay of the strongest channel path. In the $t$-th iteration this estimate is denoted by ${\widehat {\widehat { \tau}}}^{(t)}$. In step $10$, this estimated channel path delay is stored in the vector of estimated path delays ${\widehat {\widehat {\boldsymbol\tau}}}$.
	In step $12$, using (\ref{MLapproximationDoppler1}) and the estimated path delay ${\widehat {\widehat { \tau}}}^{(t)}$, we estimate the Doppler shift of the strongest channel path. Next, in step $13$, this estimated Doppler shift is stored in the vector of the estimated Doppler shifts ${\widehat {\widehat {\boldsymbol\nu}}}$. In step $14$, the complex channel path gain of the strongest channel path is estimated using (\ref{hestimate2}). In the $t$-th iteration this complex channel path gain is denoted by ${\widehat {\widehat  h} }^{(t)}$. In step $16$, we reconstruct the DD domain vector received from the strongest channel path (i.e. ${\widehat {\widehat { h}}}^{(t)} {\boldsymbol a}(\widehat{ {\widehat \tau}}^{(t)}, \widehat{ {\widehat  \nu}}^{(t)})$) and subtract this vector from the residual received vector of the previous iteration (i.e. ${\widehat {\bf x}_p}^{(t)}$). After this subtraction we get the new residual received vector ${\widehat {\bf x}_p}^{(t+1)}$. The algorithm terminates if the maximum number of allowed iterations (i.e. $T_{max}$) is reached or the difference between the normalized energy of the residual received vector in the current iteration and that in the previous iteration is less than a pre-determined threshold $\epsilon$. 

	The complexity of the proposed TSE method is $O(P m_{\tau}  M_{\tau}) + O(P n_{\nu} N_{\nu}) +  O(PM_{\tau} N_{\nu})$ as discussed in the following.
	The complexity of step $4$ and step $5$ of each iteration is $O(M_{\tau} N_{\nu})$ since most of the energy of the pilot signal is received only over an interval of $M_{\tau}$ DDREs along the delay domain
	and $N_{\nu}$ DDREs along the Doppler domain. 
	We show that the complexity of delay estimation in step $9$ of each iteration is $O(  m_{\tau}M_{\tau})$. This is because, firstly there are $(2 \lfloor m_{\tau}/2 \rfloor + 1)$ different delay values in the set $\Lambda^{(l^{(t)}-l_p)}_{\tau}$ for which the inner product ${\widehat {\bf a}}_{_{k^{(t)}+1}}({ \tau, \frac{( {k}^{(t)} - k_p)}{NT}})^H  {\widehat {\bf d}_{_{k^{(t)}+1}}}  $ needs to be computed. Further, the complexity of computing the inner product for a given delay value $\tau \in \Lambda^{(l^{(t)}-l_p)}_{\tau}$ is
	$O(M_{\tau})$ since the number of elements of the vector ${\widehat {\bf a}}_{_{k^{(t)}+1}}({ \tau, \frac{( {k}^{(t)} - k_p)}{NT}})$ having significant energy is $O(M_{\tau})$ where $M_{\tau} = \lceil M \Delta f \tau_{max} \rceil +1$.
	Similarly, the complexity of the Doppler estimation in step $12$ is $O(  n_{\nu}N_{\nu})$ since the number of delay values in the set $\Lambda^{(k^{(t)}-k_p)}_{\nu}$ is $(2 \lfloor n_{\nu}/2 \rfloor + 1)$ and the number of significant energy elements of the vector ${\widehat {\bf b}}_{l^{(t)} +1}({ {\widehat {\widehat {\tau}}}^{(t)}, \nu})$ is $O(N_{\nu})$ ($N_{\nu} = 2 \lceil \nu_{max} NT \rceil +1$).   
	Since there are roughly $O(P)$ number of iterations where $P$ is the number of channel paths, the total complexity of the proposed TSE method is $O(P m_{\tau}  M_{\tau}) + O(P n_{\nu} N_{\nu}) + O(PM_{\tau}N_{\nu})$. 
	The complexity of the TSE method is less than that of the M-MLE method, due to the fact that in the TSE method, delay and Doppler shift of each path is estimated through separate single-dimensional optimization, whereas in the M-MLE method they are jointly estimated.		
		\subsection{Comparison of Channel Estimation Complexity with Impulse/OMP/SBL Methods}
		\label{subseccmplx}
		\begin{table}
		\caption{Complexity of OTFS Channel Estimation Methods}
		\label{Complexity}
		\centering
		\begin{tabular}
			{ | c|| c| c| }
			\hline
			\multicolumn{3}{|c|}{Complexity Comparision} \\
			\hline
			Method & Matrix & Complexity\\
			 & Inversion & \\
			\hline
			Impulse \cite{EmChEst} & - &  ${ O}(M_{\tau}N_{\nu})$ \\
			\hline
			OMP \cite{OMPEsti} & $ P \times P$ &  ${ O}(P^3) + O(P^2 MN) $ \\
			& &  $ + O(PM^2N^2)$ \\
			\hline
			SBL  \cite{sparse1} & $M_{\tau}N_{\nu} \times M_{\tau}N_{\nu}$ &  ${ O}(n^{2}_{\nu} m^{2}_{\tau} N^{3}_{\nu} M^{3}_{\tau})$ \\
			\hline
			M-MLE & - &  ${ O}(P(m_{\tau}M_{\tau} n_{\nu}N_{\nu}))$ \\
			\hline
			TSE & - &  ${ O}(P m_{\tau}M_{\tau}) + O(P n_{\nu}N_{\nu})$  \\
			&  & $ + { O}(P M_{\tau}N_{\nu}) $ \\
			\hline
		\end{tabular}
	\end{table}
	In Table-\ref{Complexity} we list the complexity of the proposed methods (M-MLE and TSE) along with that of the other OTFS channel estimation methods in prior literature.
	We specifically consider the Impulse method, the OMP method and the SBL method proposed in \cite{EmChEst}, \cite{OMPEsti} and \cite{sparse1} respectively.
	In the Impulse method proposed in  \cite{EmChEst}, the channel estimation pilot signal consists of energy transmitted on a single DDRE in the DD domain (i.e., an impulse in the DD domain).
	The DD domain channel is then estimated at the receiver from the symbols received on the DDREs around the pilot DDRE (i.e., the DDRE where the pilot signal energy was transmitted).
	As the received pilot energy is localized around the pilot DDRE in a region of width $M_{\tau}$ DDREs along the delay domain and $N_{\nu}$ DDREs along the Doppler domain,
	the total complexity of this impulse method is $O(M_{\tau}N_{\nu})$. In the OMP method proposed in \cite{EmChEst}, in each iteration, the product of a $MN \times MN$ matrix with a $MN \times 1$ vector is required
	which has a per-iteration complexity of $O(M^2N^2)$ (which is high for large $(M,N)$). In addition, inversion of a $P \times P$ matrix is also required in the OMP method.  
         In the SBL method proposed in \cite{sparse1}, the inversion of a $M_{\tau}N_{\nu} \times M_{\tau}N_{\nu}$ matrix is required which has a complexity of $O(M^{3}_{\tau}N^{3}_{\nu} )$ (which is high
         for large $(M,N)$).
         
         From Table-\ref{Complexity}, we observe that the complexity of the proposed methods (M-MLE and TSE) is significantly \emph{smaller} than that of
         the OMP and the SBL methods when $(M,N)$ is large. Also, although the complexity of the Impulse based method is smaller than that of the proposed methods,
         its NMSE and error-rate performance are significantly inferior to that of the proposed TSE and M-MLE methods (see Section \ref{simsec1}).
         
         Regarding the pilot overhead required for channel estimation, it is observed that most of these methods use impulse like pilots in the DD domain for which
         only a region of $O(M_{\tau}N_{\nu})$ DDREs out of the total $MN$ DDREs needs to be reserved/dedicated in the OTFS frame. The remaining DDREs
         could be used for transmission of information. Additionally, since the effective DD domain channel varies slowly, frequent DD domain channel estimation is
         generally not required, i.e., we need not send pilots in each OTFS frame.  

	\section{Numerical Results}
	\label{simsec1}
	In this section, we present the results of numerical studies carried out by us to assess the quality of the estimate of the effective DD domain channel obtained using our proposed methods (M-MLE and TSE) when compared
	to other channel estimation methods known in prior literature (i.e., Impulse method \cite{EmChEst}, OMP method \cite{OMPEsti} and the SBL method \cite{sparse1}).
	A comparison of the channel estimation complexity of the proposed M-MLE and TSE methods with that of the Impulse, OMP and SBL methods is given in Table-\ref{Complexity}.
	
	For the proposed methods (M-MLE and TSE) and the Impulse and SBL methods, we consider rectangular transmit and receive pulses.
	The SBL method considered in this section is the ``$1$D Off-grid SBL" method proposed in \cite{sparse1}. In \cite{sparse1}, this method has been proposed
	for ideal transmit and receive pulses which are not realizable in practice. Therefore, for the comparison here, we have adapted this method for practical rectangular pulses.
	The OMP method presented in \cite{OMPEsti} assumes ideal transmit and receive pulses and it is difficult to adapt it for non-ideal pulses.\footnote{\footnotesize{In \cite{OMPEsti}, due to ideal pulses, the effective DD domain channel is a 2-D convolution in the DD domain
	which reduces the maximum possible number of DD domain channel coefficients from $M^2N^2$ in case of non-ideal/practical transmit and receive pulses to only $MN$ with ideal pulses. As the method proposed
	in  \cite{OMPEsti} is based on the 2-D convolutive DD domain channel model, it is difficult to adapt it for non-ideal pulses.}} Therefore the presented OMP simulation results
	 serve as an upper bound on the best possible performance of this OMP method with practical non-ideal pulses.

	 We consider OTFS modulation with $\Delta f = 30$ KHz and $T = 1/\Delta f = 33. 33 \mu s$.
	For the numerical studies we consider the wireless channel between an aircraft and the ground station during the aircraft's arrival.
	The channel model for this aircraft arrival scenario is based on the model in \cite{HaarChannel}.
	We consider $P = 5$ paths, where the delay of the first path i.e., direct line-of-sight path (LOS) path is $\tau_1 = 0$ and the delay of all other paths is distributed uniformly in $(0 \,,\, 7 \mu s ]$. A Rice-factor of $K = 15$ dB is considered and therefore the fixed absolute squared value of the channel gain of the LOS path is $\vert h_1 \vert^2 = K/(K+1)$.  The mean squared value of the channel gains of the other paths are modelled using an exponential power delay profile given by equation $(7)$ in \cite{HaarChannel} with $\tau_{\mbox{\tiny{slope}}} = 1 \mu s$. The channel gains corresponding to the other paths (i.e., $h_i, i=2,\cdots, P$) are modelled as i.i.d.
	Rayleigh faded and their mean squared values are normalized so that the sum of the mean squared values of the channel gains of these other paths is $1/(K+1)$. The Doppler shift of the
	direct LOS path is taken to be $\nu_1 = \nu_{max}$ while for the other paths $\nu_i = \nu_{max} \cos(\theta_i)$ where $\theta_i$ is distributed uniformly in the interval $(0 \,,\, 2 \pi]$.
        We consider an arrival speed of $100$ m/s (i.e., $360$ Km/hr) and a carrier frequency of $f_c = 5.1$ GHz which corresponds to $\nu_{max} = 1700$ Hz.

         We firstly compare the normalized mean square error (NMSE) of the effective DD domain channel estimate obtained using the proposed method with that of the NMSE of other estimators
         known in prior literature.
         For any estimator, the corresponding NMSE is given by $\mathbb{E} \left[ \frac{\Vert {\bf G} - {\widehat { {\bf G}}} \Vert^2_F}{\Vert {\bf G} \Vert^2_F} \right]$ where ${\bf G}$
         denotes the actual effective DD domain channel matrix and ${\widehat { {\bf G}}}$ denotes the estimated matrix. Also, in the following, pilot signal-to-noise ratio (PSNR) is defined as the ratio of the
	average power of the transmitted time-domain pilot signal to the AWGN power at the receiver. For example, for the proposed methods,
	$E_p$ is the total energy of the transmitted time-domain pilot signal of duration $NT$ seconds, and therefore
	its power is $E_p/(NT)$ (see (\ref{Pilot1}) and the sentence after it). Since the communication bandwidth is $M \Delta f$ Hz and the noise power
	spectral density is $N_0$, the total noise power at the receiver is $M \Delta f N_0$ and hence the PSNR is the ratio of $E_p/(NT)$ to $M \Delta N_0$ which is
	$E_p/(M N N_0)$. Since the channel path gains are normalized (i.e., $\sum\limits_{i=1}^P | h_i |^2 = 1$), the ratio of the average received pilot power
	to the AWGN power is also $E_p/(M N N_0)$.
         
\begin{figure}[h!]
	\centering
	\includegraphics[width=0.9\linewidth]{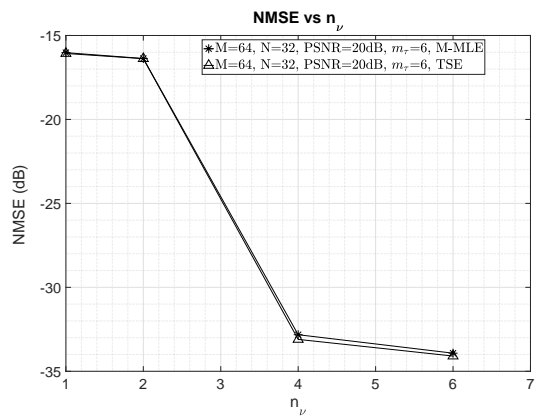}
	\caption{NMSE vs. $n_{\nu}$. Fixed $M = 64, N=32$, $m_{\tau} = 6$, $\mbox{\small{PSNR}}=20$ dB.}
	\label{fig2}
\end{figure}

\begin{figure}[h!]
	\centering
	\includegraphics[width=0.9\linewidth]{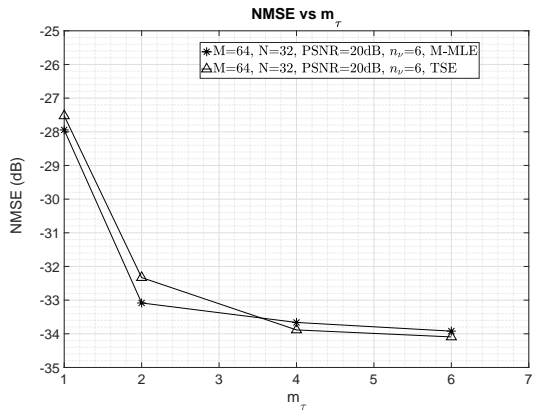}
	\caption{NMSE vs. $m_{\tau}$. Fixed $M = 64, N=32$, $n_{\nu} = 6$, $\mbox{\small{PSNR}}=20$ dB.}
	\label{fig3}
\end{figure}

         For the proposed M-MLE and TSE methods, through Fig.~\ref{fig2} and Fig.~\ref{fig3} we show that it suffices to consider $m_{\tau} =   n_{\nu} = 6$.
         For the proposed methods we consider the maximum number of allowed iterations to be $T_{max} = 15$ for all results presented in this section.
         In Fig.~\ref{fig2}, we plot the NMSE of the proposed methods as a function of increasing $n_{\nu}$, for a fixed $M = 64, N = 32$, PSNR $=20$ dB and $m_{\tau} = 6$.
         It is observed that with increasing $n_{\nu}$, the NMSE decreases due to the improvement in the refinement of the possible Doppler shift values. 
         However, with further increase in $n_{\nu}$, the amount of decrease in NMSE is insignificant for $n_{\nu} \geq 4$.
         Since a large $n_{\nu}$ also implies higher estimation complexity, we therefore consider $n_{\nu} = 6$ for the subsequent numerical studies reported in this section
         for both the proposed methods.
         
         In Fig.~\ref{fig3} we plot the NMSE of the proposed methods as a function of increasing $m_{\tau}$, for a fixed $M = 64, N = 32$, PSNR $=20$ dB and $n_{\nu} = 6$.
         It is observed that with increasing $m_{\tau}$, as expected the NMSE decreases although the amount of decrease is insignificant for $m_{\tau} \geq 4$.
         Since, a large $m_{\tau}$ also implies higher complexity, we therefore consider $m_{\tau} = 6$ for the subsequent numerical studies reported in this section
         for both the proposed methods.     

	
\begin{figure}[h!]
		\includegraphics[width=0.95\linewidth]{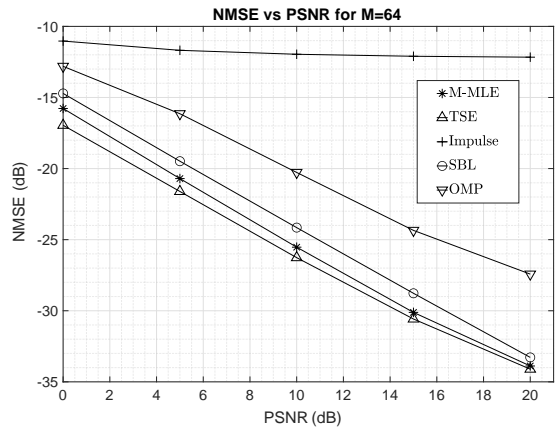}
		\caption{NMSE vs. PSNR. $M=64$, $N=32$.}
		\label{fig4}
	\end{figure}

	In Fig.~\ref{fig4} we plot the NMSE of the proposed methods and also that of the other methods known in prior literature, as a
	function of increasing PSNR for $M = 64, N = 32$. In Fig.~\ref{fig4} it is observed that the proposed methods achieve better NMSE performance than the other methods.
	Although the NMSE performance of the SBL method is close to that of the proposed method, the SBL method has a significantly higher complexity compared to
	the proposed methods as it requires the inversion of a large $M_\tau N_\nu \times M_\tau N_\nu$ matrix whereas the proposed methods do not require
	any matrix inversion (note that for $M = 64, N = 32$ and $\tau_{max} = 7 \mu s, \nu_{max} = 1700$ Hz, we have $M_\tau = \lceil M \Delta f \tau_{max} \rceil +1 = 15, N_{\nu} = 2 \lceil \nu_{max} NT \rceil +1=5$
	and therefore the SBL method requires the inversion of a $75 \times 75$ matrix).
	
	In Fig.~\ref{fig4}, the NMSE performance of the Impulse based method is poor as it is based only on the 
	estimation of the channel gains of the effective DD domain channel matrix and does not exploit the underlying system model (i.e., it does not exploit the dependence of the effective DD domain channel matrix
	on the channel path delay and Doppler shifts). In \cite{OMPEsti}, the OMP method is shown to achieve
	accurate estimation of the effective DD domain channel matrix for a channel scenario with integer delay and Doppler shifts (i.e., where the multi-path delay and Doppler
	shifts are integer multiples of the delay domain resolution (i.e., $T/M$) and the Doppler domain resolution (i.e., $\Delta f/N$) respectively). However,
	in our numerical studies we consider non-integer delay and Doppler shifts, and this is why in Fig.~\ref{fig4} the performance of the OMP method is inferior to that of the proposed
	methods. One reason is that the channel estimation accuracy of the OMP based method relies on the sparsity of the effective DD domain channel. However, in the presence of non-integer delay and Doppler shifts, the effective DD domain channel is not as sparse as that in an ideal scenario with integer delay and Doppler shifts.
	
\begin{figure}[h!]
	\includegraphics[width=0.95\linewidth]{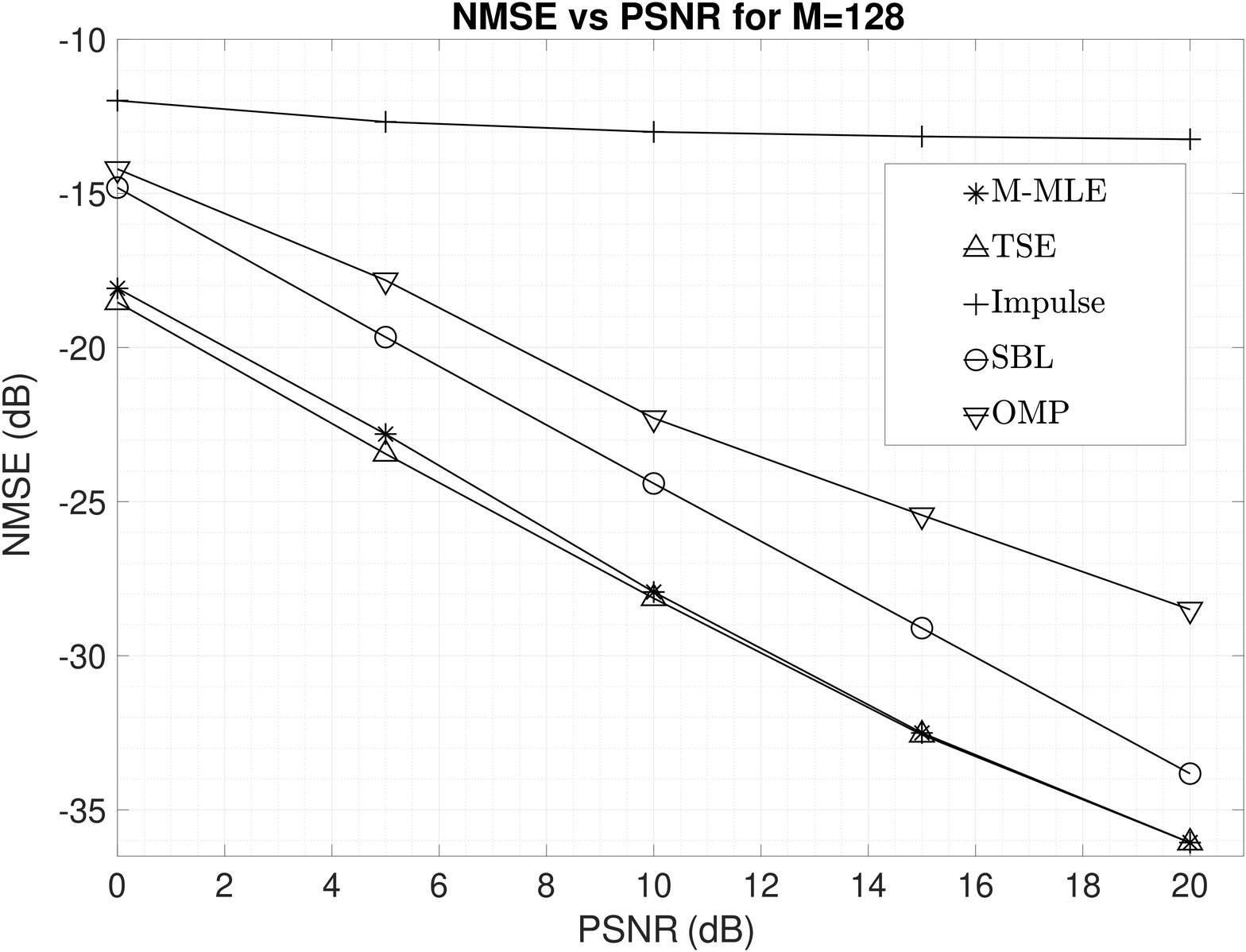}
	\caption{NMSE vs. PSNR. $M=128$, $N=32$.}
	\label{fig5}
     \end{figure}

	In Fig.~\ref{fig5}, just as in Fig.~\ref{fig4} we again compare the NMSE performance of all methods versus PSNR, but for a higher $M = 128$. We observe that at high PSNR,
	when compared with Fig.~\ref{fig4}, the NMSE performance of all methods improve. This improvement appears to be due to the improvement in the delay domain resolution $T/M$ when $M$ is doubled from
	$M=64$ in Fig.~\ref{fig4} to $M=128$ in Fig.~\ref{fig5}. For a fixed PSNR of $20$ dB, the reduction in the NMSE (when compared to Fig.~\ref{fig4}) is roughly $2$ dB
	for both the proposed methods whereas it is only about $1.0$ dB for both the SBL and the OMP methods.  
	               
\begin{figure}[h!]
		\includegraphics[width=0.95\linewidth]{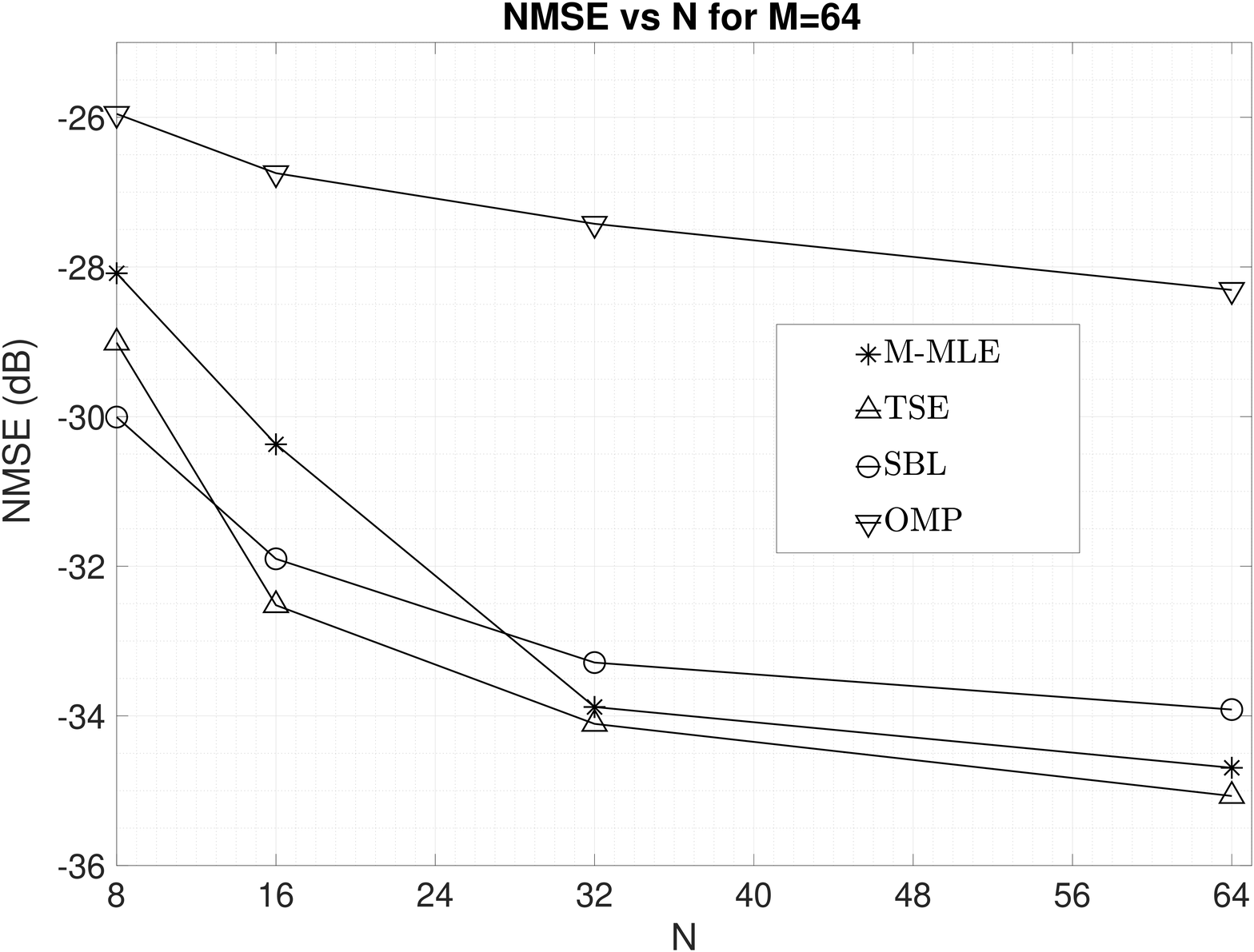}
		\caption{NMSE vs. $N$. $M=64$, $\mbox{\small{PSNR}} = 20$ dB.}
		\label{fig6}
	\end{figure}

	In Fig.~\ref{fig6} we plot the NMSE performance as a function of increasing $N$ for a fixed $M=64$ and a fixed $\mbox{\small{PSNR}} = 20$ dB.
	For all methods, it is observed that the NMSE reduces with increasing $N$. This reduction is primarily due to the improvement in the
	Doppler domain resolution $\Delta f/N$ with increasing $N$. It is also observed that, although for small $N \leq 16$ (i.e., insufficient Doppler domain resolution) the NMSE of the proposed methods (M-MLE and TSE) is inferior to that of the SBL method, for sufficiently large $N$ ($N \geq 32$), the NMSE of the proposed methods is better than that of SBL and the other considered methods. It is known that the robustness of OTFS modulation to channel path delay and Doppler shifts is primarily due to the joint demodulation of all DD domain information symbols which is practically feasible only when the effective DD domain channel is sparse and for which the delay and Doppler domain resolution should be sufficiently large \cite{HadaniOTFS2, SKM4}. Hence, $(M,N)$ would anyways be large in OTFS based systems and for such practical scenarios the estimation accuracy (i.e., NMSE) of the proposed methods is observed to be better than that of the other estimation methods (see Fig.~\ref{fig4}, Fig.~\ref{fig5} and Fig.~\ref{fig6}).    
	
\begin{figure}[h!]
		\includegraphics[width=0.95\linewidth]{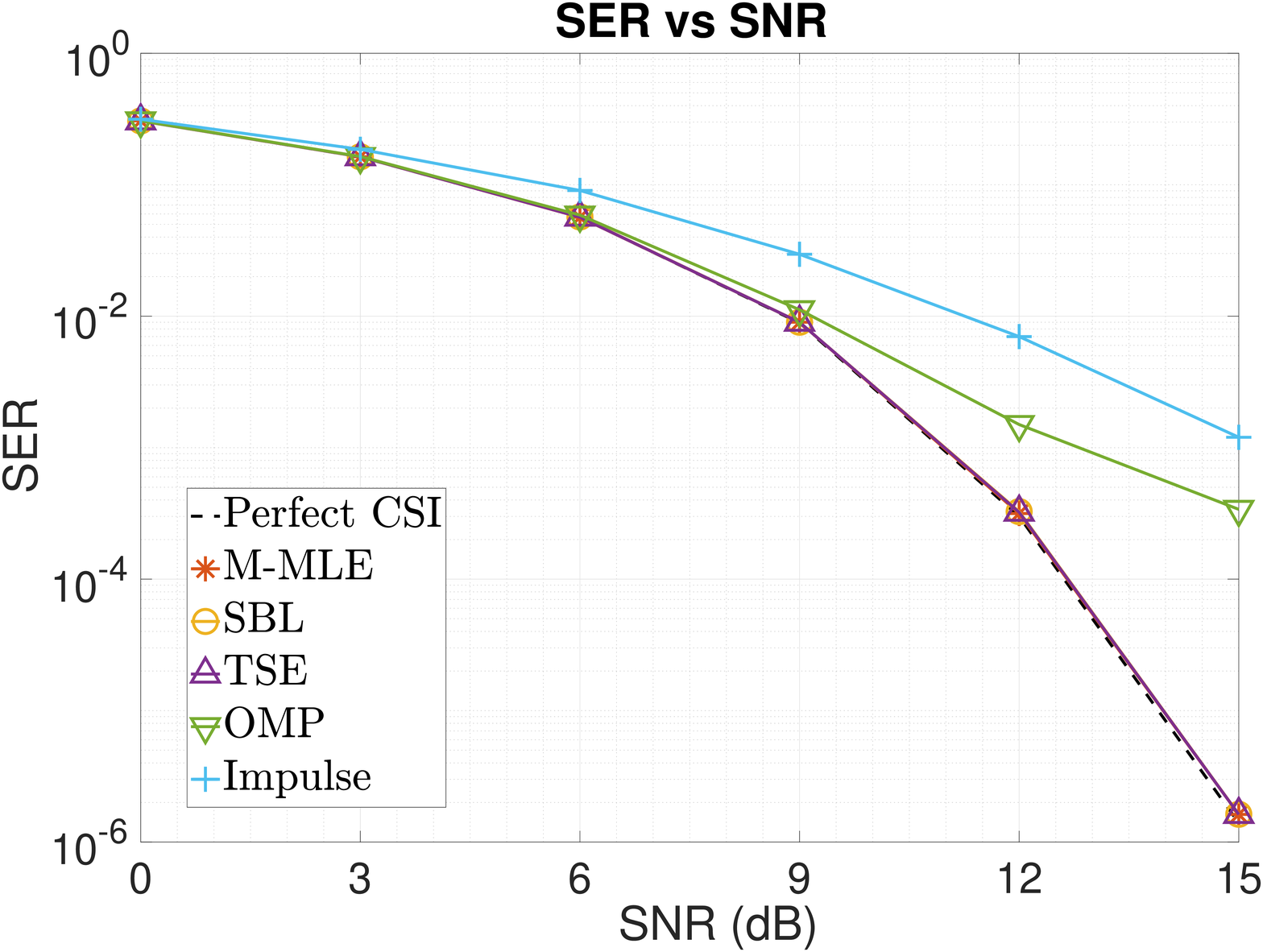}
		\caption{SER vs. SNR. $M=64$, $N=32$, $\mbox{\small{PSNR}}=15$ dB.}
		\label{fig7}
	        \end{figure}

\begin{figure}[h!]
		\vspace{-4mm}
		\includegraphics[width=0.95\linewidth]{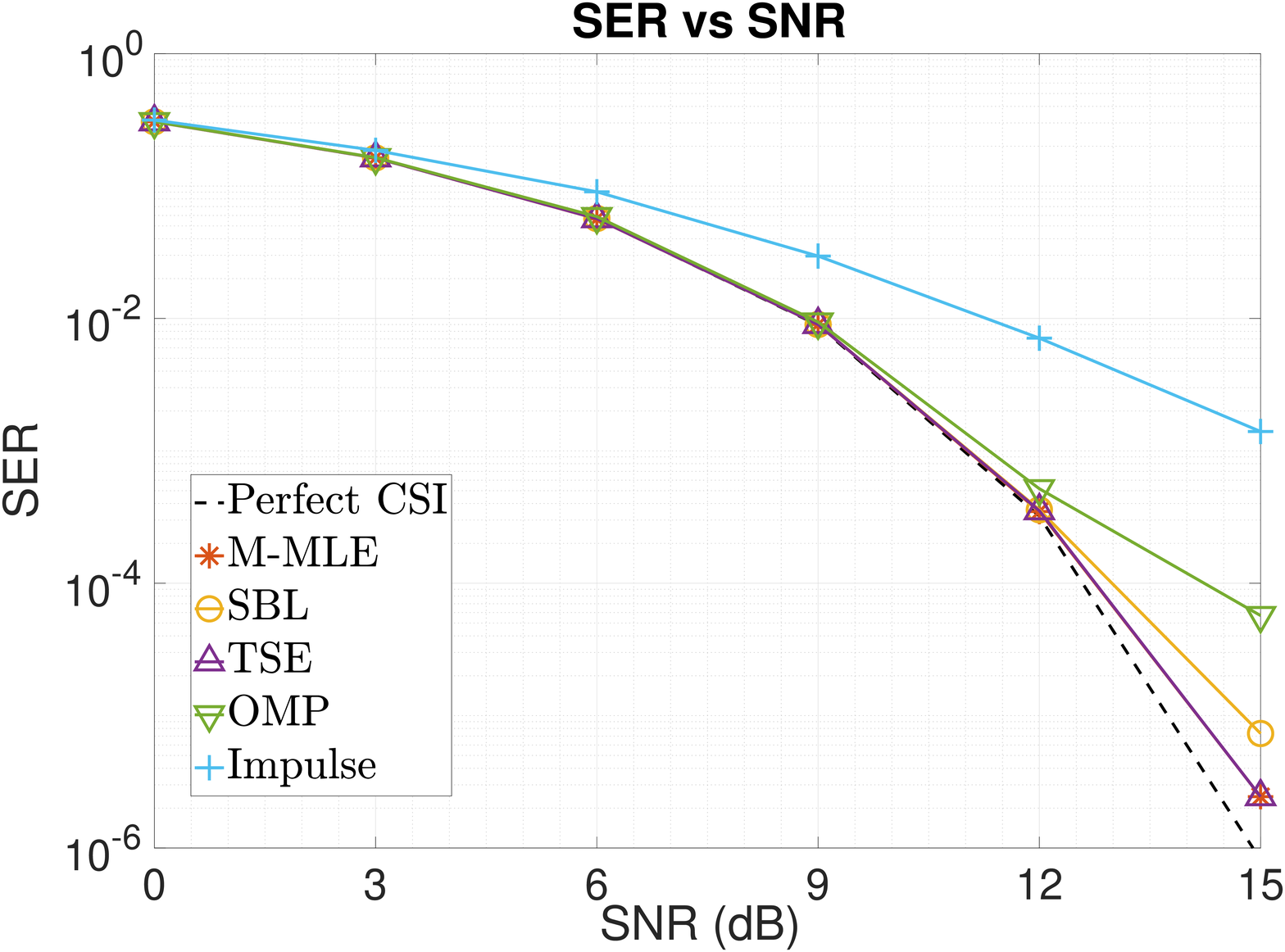}
		\caption{SER vs. SNR. $M=128$, $N=32$, $\mbox{\small{PSNR}}=15$ dB.}
		\label{fig8}
	        \end{figure}

	In Fig.~\ref{fig7} we plot the uncoded $4$-QAM symbol error rate (SER) performance of the considered methods as a function of increasing signal-to-noise ratio (SNR), when a DD domain message passing (MP) detector (see \cite{channel})
	is used at the receiver. MP detection is performed with the estimated effective DD domain channel matrix ${\widehat { {\bf G}}}$.
	We have also plotted the SER performance when the effective DD domain channel matrix is known perfectly at the receiver (see ``Perfect CSI" in the legend of Fig.~\ref{fig7}).
	We consider a fixed $M = 64, N=32$, $\mbox{\small{PSNR}} = 15$ dB. SNR is the ratio of the average total received signal power (i.e., of the information carrying time-domain OTFS modulated signal)
	to the AWGN power at the receiver. It is observed that the SER performance with the proposed estimation methods (M-MLE and TSE) and the SBL method is same as the SER
	performance with perfect channel estimates. The SER performance with the OMP and the Impulse estimation methods is however inferior to that achieved with the proposed and the SBL method.
	This is expected as in Fig.~\ref{fig4}, Fig.~\ref{fig5}, and Fig.~\ref{fig6} we have seen that the NMSE performance of the proposed methods is the best among all considered methods.
	In Fig.~\ref{fig8}, we plot the uncoded $4$-QAM SER for all the considered methods for the same setting as in Fig.~\ref{fig7}, but with a higher $M = 128$.
	With $M = 128$, the SER performance of the proposed methods is better than that of the other methods and is close to the ideal SER performance with perfect CSI.
	Note that with $M=128$, at high SNR the SER performance of the proposed methods is slightly better than that of the SBL method.
	
	From the numerical results presented in this section and the complexity comparison in Section \ref{subseccmplx}, it can be concluded that with sufficiently large $(M,N)$, the
	proposed estimation methods achieve the best NMSE and SER performance (among the considered methods) at low complexity.
	The closeness of the SER achieved by the proposed methods to the SER achieved with perfect CSI reveals the effectiveness of the proposed channel estimation methods.  

	\section{Conclusion}
	In this paper, we have proposed two low-complexity channel estimation methods (M-MLE and TSE) for OTFS based systems.
	These estimation methods are based on the observation that, due to the fine delay and Doppler domain resolution in OTFS based systems, the high complexity joint ML estimation of the multi-path channel gain, delay and Doppler shifts can be decoupled into separate ML estimation of the gain, delay and Doppler shift of each path. 
	The proposed methods do not require matrix inversion and have lower complexity than that of other methods known to achieve good channel estimation accuracy (i.e., OMP and SBL). Through simulations, we also show that the proposed
	methods achieve better channel estimation accuracy than other known methods when the delay and Doppler domain resolution is sufficiently fine.    
		
	\appendices
	\section{Asymptotic Orthogonality of Columns of ${\bf A}\left( {\boldsymbol \tau}, {\boldsymbol \nu} \right)$}
	\label{appendixA}
	Let us consider two columns of ${\bf A}\left( {\boldsymbol \tau}, {\boldsymbol \nu} \right)$ corresponding to two channel paths having delay and Doppler shifts $(\tau_1, \nu_1)$
	and $(\tau_2, \nu_2)$ respectively. These columns are then given by ${\bf a}(\tau_1, \nu_1)$ and ${\bf a}(\tau_2, \nu_2)$ respectively where ${\bf a}(\tau,\nu)$ is defined in the paragraph after
	(\ref{rxpilotvec}). The expression for $\vert {\bf a}(\tau_1, \nu_1)^H {\bf a}(\tau_2, \nu_2) \vert$ is given by (\ref{eqnorthogonal1}) (see top of next page).
	{
	\begin{figure*}
	\vspace{-8mm}
	{\small
	\begin{eqnarray}
	\label{eqnorthogonal1}
	\left\vert {\bf a}(\tau_1, \nu_1)^H{\bf a}(\tau_2, \nu_2) \right\vert & = & M N E_p \left\vert \sum\limits_{l'=0}^{M-1}\sum\limits_{k'=0}^{N-1} b_{q,1,l',k'}^*  \, b_{q,2,l',k'}  \right\vert \nonumber \\
	& \mya & M N E_p \left\vert \sum\limits_{l'=0}^{M-1}\sum\limits_{k'=0}^{N-1} e^{j 2 \pi (\nu_1 \tau_1 - \nu_2 \tau_2)}  \left[   \frac{1}{N^2} \sum\limits_{n_1 = 0}^{N-1} \sum\limits_{n_2 = 0}^{N-1}  e^{j 2 \pi (n_1 - n_2) \frac{k' - k_p}{N} } e^{-j 2 \pi (n_1 \nu_1 - n_2 \nu_2) T} \right]  g_1^*[l'] g_2[l']  \right\vert \nonumber \\
	& \myb &  M N E_p \left\vert  \sum\limits_{l'=0}^{M-1}   g_1^*[l'] g_2[l']  \right\vert  \,  \left\vert    \frac{1}{N} \sum\limits_{n_1 = 0}^{N-1}  \sum\limits_{n_2 = 0}^{N-1}   e^{- j 2 \pi (n_1 - n_2) \frac{k_p}{N} } e^{-j 2 \pi (n_1 \nu_1 - n_2 \nu_2) T}   \left[ \underbrace{\frac{1}{N} \sum\limits_{k'=0}^{N-1}  e^{j 2 \pi (n_1 - n_2) \frac{k'}{N} } }_{= \delta[n_1 - n_2]}\right] \right\vert  \nonumber \\
	&  \myc  &  M N E_p  \left\vert  \sum\limits_{l'=0}^{M-1}   g_1^*[l'] g_2[l']  \right\vert  \,  \left\vert    \frac{1}{N} \sum\limits_{n = 0}^{N-1}   e^{-j 2 \pi n (\nu_1 - \nu_2) T}   \right\vert  \, = \,  M N E_p \left\vert  \sum\limits_{l'=0}^{M-1}   g_1^*[l'] g_2[l']  \right\vert  \,  \left\vert \frac{\sin\left( \pi N (\nu_1 - \nu_2) T\right)}{N \sin\left( \pi (\nu_1 - \nu_2) T  \right)}  \right\vert \nonumber \\
	g_i[l'] & \Define &  \left[\frac{1}{M} \sum\limits_{m=0}^{M-1} e^{j 2 \pi \frac{m}{M} \left( l' - l_p -M \tau_i \Delta f\right)} f_{\tau_i,\nu_i,k_p,l'}(m) \right] \,\,,\,\, i=1,2.
	\end{eqnarray}
			{\vspace{-3mm}
		\begin{eqnarray*}
		\hline
		\end{eqnarray*}}
		\vspace{-5mm}
	\normalsize}
	\end{figure*}}
	Step (a) of (\ref{eqnorthogonal1}) follows from the fact that $q = (k_p M + l_p +1)$ (see paragraph after (\ref{rxpilotvec})). 
	The definition of $g_i[l']$ follows from the expression of $b_{q,i,l',k'}$ in (\ref{apikleqn}) for $q= (k_p M + l_p +1)$.
	In Step (b) we separate the delay domain term (summation index $l'$) and the Doppler domain term (summation indices $n_1,n_2,k'$). In the Doppler domain term we change the order of summation (i.e.,
	the summation over $k'$ is now the innermost summation). This innermost summation is $\delta[n_1 - n_2]$ where, $\delta[n] = 1$ if $n=0$ and is zero otherwise. Therefore, the innermost summation over $k'$ is non-zero
	only when $n_1 = n_2$. Step (c) then follows from this fact. 
	As  (\ref{eqnorthogonal1}) is valid for any $(\tau_1,\nu_1)$ and any $(\tau_2,\nu_2)$, for the special case of $\tau_2=\tau_1$ and $\nu_2 = \nu_1$ we get, $\Vert {\bf a}(\tau_1, \nu_1) \Vert_2^2 = {\bf a}(\tau_1, \nu_1) ^H {\bf a}(\tau_1, \nu_1) =  M N E_p \sum\limits_{l'=0}^{M-1}   \vert g_1[l'] \vert^2$. Similarly, we also get  $\Vert {\bf a}(\tau_2, \nu_2) \Vert_2^2 =  M N E_p \sum\limits_{l'=0}^{M-1}   \vert g_2[l'] \vert^2$. For $\nu_1 \ne \nu_2$, using (\ref{eqnorthogonal1}) we get
	
	{\vspace{-4mm}
	\small
	\begin{eqnarray}
	\label{eqnorthogonal2}
	\lim_{N \rightarrow \infty} \frac{\left\vert {\bf a}(\tau_1, \nu_1)^H{\bf a}(\tau_2, \nu_2) \right\vert}{\Vert {\bf a}(\tau_1, \nu_1) \Vert_2 \,  \Vert {\bf a}(\tau_2, \nu_2) \Vert_2} & \hspace{-3mm} = &  \hspace{-3mm} \frac{\left\vert  \sum\limits_{l'=0}^{M-1}   g_1^*[l'] g_2[l']  \right\vert }{\sqrt{\sum\limits_{l'=0}^{M-1}   \vert g_1[l'] \vert^2} \sqrt{\sum\limits_{l'=0}^{M-1}   \vert g_2[l'] \vert^2}} \nonumber \\
	& & \hspace{-3mm} \,  \lim_{N \rightarrow \infty}  \left\vert \frac{\sin\left( \pi N (\nu_1 - \nu_2) T\right)}{N \sin\left( \pi (\nu_1 - \nu_2) T  \right)}  \right\vert  \nonumber \\
	& = & 0
	\end{eqnarray} 
	\normalsize}which shows that ${\bf a}(\tau_1, \nu_1)$ and ${\bf a}(\tau_2, \nu_2)$ are asymptotically orthogonal if $\nu_1 \ne \nu_2$.
	In (\ref{eqnorthogonal2}), the last step follows from the fact that
	
	{\vspace{-4mm}
	\small
	\begin{eqnarray}
	\label{eqnorthogonal3}
	\lim_{N \rightarrow \infty} \left\vert \frac{\sin\left( \pi N (\nu_1 - \nu_2) T\right)}{N \sin\left( \pi (\nu_1 - \nu_2) T  \right)}  \right\vert & \hspace{-3mm} =  & \hspace{-3mm} \frac{\vert \pi (\nu_1 - \nu_2) T \vert }{\vert \sin\left( \pi (\nu_1 - \nu_2) T  \right) \vert }  \nonumber \\
	&   &  \hspace{-3mm}  \lim_{N \rightarrow \infty} \left\vert \frac{\sin\left( \pi N (\nu_1 - \nu_2) T\right)}{\pi N (\nu_1 - \nu_2) T} \right\vert \nonumber \\
	& \hspace{-49mm}  = &  \hspace{-26mm} \frac{\vert \pi (\nu_1 - \nu_2) T \vert }{\vert \sin\left( \pi (\nu_1 - \nu_2) T  \right) \vert }  \,  \lim_{N \rightarrow \infty} \left\vert \mbox{\small{sinc}}\left( N (\nu_1 - \nu_2) T \right) \right\vert   \nonumber \\
	& \hspace{-49mm}  = &  \hspace{-26mm} 0
	\end{eqnarray}
	\normalsize}since $\lim_{x \rightarrow \infty} \mbox{\small{sinc}}(x) = 0$. Since $\vert \mbox{\small{sinc}}(x) \vert$ is small for $\vert x \vert \gg 1$, from (\ref{eqnorthogonal3})
	it follows that the two columns ${\bf a}(\tau_1, \nu_1)$ and ${\bf a}(\tau_2, \nu_2)$ are almost orthogonal if $\vert N (\nu_1 - \nu_2) T\vert \gg 1$ i.e., if $\frac{1}{NT} \ll \vert \nu_1  - \nu_2 \vert$.

\end{document}